\documentclass[9pt,twocolumn,twoside]{opticajnl}
\journal{opticajournal} % use for journal or Optica Open submissions

\setboolean{shortarticle}{false}
\usepackage{lineno}
\usepackage{graphicx}
\usepackage{epsfig}
\usepackage{subfigure}
\usepackage{rotating}
\usepackage{mathrsfs}
\usepackage{amsfonts}
\usepackage{amsmath}
\usepackage{amssymb}
\usepackage{placeins}
\usepackage{bm}

\usepackage{pgfplots}
\usepackage{tikz}
\usetikzlibrary{arrows,decorations.markings}
\pgfplotsset{compat=1.14}

\usepackage{soul}
\usepackage{fixmath}
\usepackage{physics}

\newcommand{\hc}{\ensuremath{\text{h.c.}}}
\newcommand{\transp}{\ensuremath{\mathrm{T}}}

\DeclareMathOperator{\diag}{diag}

\newtheorem{property}{Property}
\newtheorem{proposition}{Proposition}
\newtheorem{lemma}{Lemma}

\newcommand\fig[1]{Fig.\ref{fig:#1}}

\usepackage{ulem}

\title{Universal quantum frequency comb measurements by spectral mode-matching}

\author[1]{Bakhao Dioum}
\author[2]{Virginia D'Auria}
\author[3,4]{Alessandro Zavatta}
\author[5]{Olivier Pfister}
\author[1,*]{Giuseppe Patera}

\affil[1]{University of Lille, CNRS, UMR 8523 - PhLAM - Physique des Lasers Atomes et Mol\'{e}cules, F-59000 Lille, France}
\affil[2]{Université Côte d'Azur, CNRS, Institut de Physique de Nice (INPHYNI), UMR 7010, 17 rue J. Lauprêtre, 06200 Nice, France}
\affil[3]{Istituto Nazionale di Ottica, Consiglio Nazionale delle Ricerche (CNR-INO), L.go E. Fermi 6, Firenze 50125, Italy.}
\affil[4]{QTI S.r.l., L.go E. Fermi 6, Firenze 50125, Italy.}
\affil[5]{Department of Physics, University of Virginia, 382 McCormick Road, Charlottesville, VA 22903, USA}

\affil[*]{giuseppe.patera@univ-lille.fr}

\begin{abstract}
The frequency comb of a multimode interferometer offers exceptional scalability potential for field-encoded quantum information. However, the staple field detection method, homodyne detection, cannot access quantum information in the whole comb because some spectral quadratures (and their asymmetries with respect to the LO) are out of reach. We present here the first general approach to make arbitrary, one-shot measurements of a multimode quantum optical source, something that is required for photonic quantum computing and is not possible when using homodyne detection with a pulse-shaped LO. This approach uses spectral mode-matching, which can be understood as interferometry with a memory effect. We derive a complete formalism and propose an implementation by microcavity arrays.
\end{abstract}
\setboolean{displaycopyright}{false}

\begin{document}

%\pacs{vvv}

%\date{\today}
\maketitle
%%%%%%%%%%%%%%%%%%%%%%%%%%%%%%%%%%%%%%%%%%%%%%%%%%%%%%%%%%%%%%%%%%%%%%%%%%%%%%%%%%%%%%%%%%%%%%%%%%%%
\section{Introduction}

Measurement plays a distinct part in quantum mechanics. As such, it serves multiple purposes in quantum information processing (QIP): it is of course essential to the characterization of quantum states but its projective nature is also a valuable tool for state preparation and is central to measurement-based  quantum computation (MBQC). Continuous-variable, a.k.a.\ field- or qumode-based, MBQC makes use of hybrid bosonic qubit encodings~\cite{Pfister2024}, such as the Gottesman-Kitaev-Preskill (GKP) encoding~\cite{Gottesman2001} which allows for universal fault-tolerant MBQC~\cite{Menicucci2014ft}.  The use of multimode quantum fields for MBQC has shown exceptional scalability potential~\cite{Chen2014,Yokoyama2013,Yoshikawa2016,Asavanant2019,Larsen2019}. In this context, both undulatory (field-resolved) and corpuscular (photon-number-resolved---PNR) measurements are mandated. The latter have recently come of age thanks to superconducting transition edge sensors~\cite{Lita2008,Eaton2023} and to superconducting nanowire detectors~\cite{Cahall2017}. In contrast, field-resolved measurements have much longer been optimal (with negligible losses) using homodyne detection (HD).\footnote{Note that hybrid detection has also been investigated that includes a weak local oscillator and PNR detection, for example in computationally efficient quantum state tomography~\cite{Wallentowitz1996,Banaszek1996,Sridhar2014a,Nehra2019,Olivares2019,Nehra2020a}.} 
%, \fig{HD}). 
In MBQC as well as in any scheme involving time-homodyne such as for heralded quantum state preparation~\cite{Lvovsky2013,melalkia_plug-and-play_2022,RaMultiNG2020}, arbitrary multimode, single-shot field measurements can be required, which faces significant challenges. 
In this paper, we present the first universal approach for overcoming these challenges and making such measurements.

The core of HD consists in matching a local oscillator (LO) field mode to the  qumode to be measured. This mode-matching implies spatial (longitudinal and transverse) wavefront matching but also temporal, or spectral, mode-matching. As is well known, any deviation from 100\% mode-matching leads to contamination of the measured qumode by fluctuations from other qumodes, e.g.\ vacuum ones. This problem, %generally 
especially insidious for spectro-temporal degrees of freedom, can be directly analyzed by decomposing the signal onto orthogonal modes with respect to the LO, or vice versa~\cite{Brecht2015}. 
\begin{figure*}[ht]
  \centering
  \includegraphics[width=\linewidth]{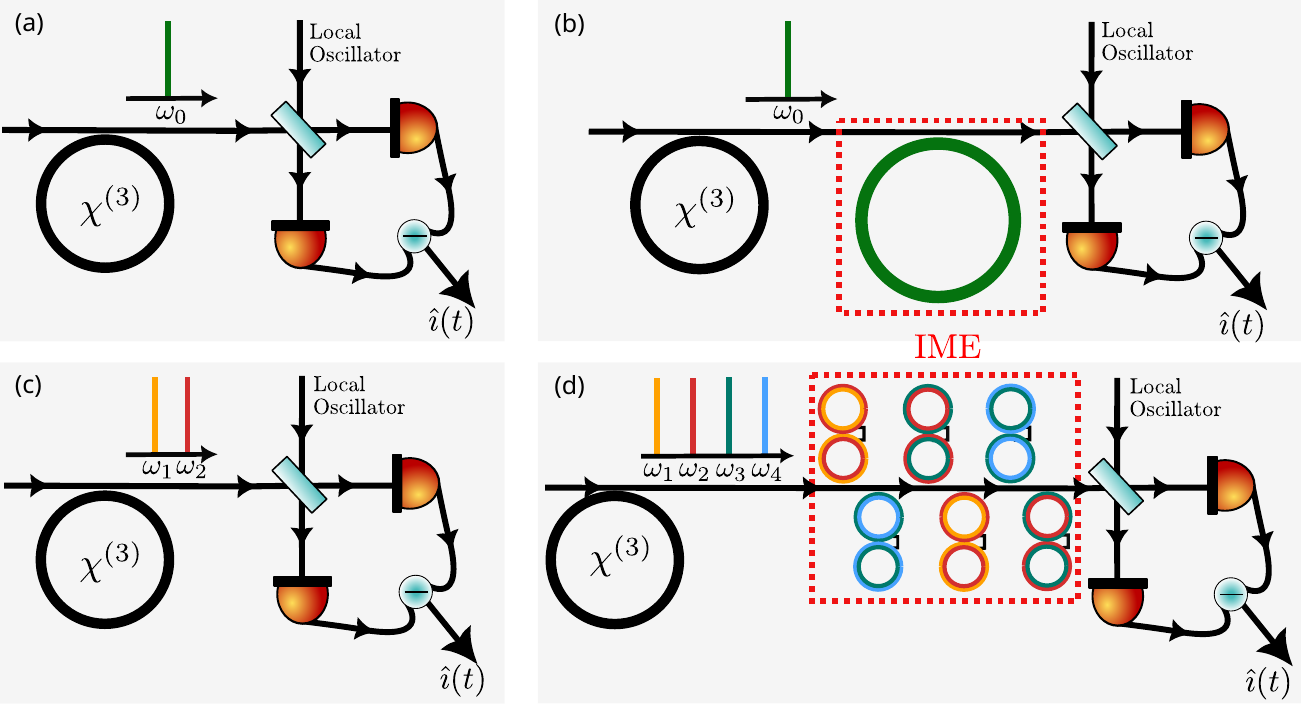}
  \caption{(a) Scheme of a standard HD for the characterization of a single-mode quantum field with carrier $\omega_0$ generated by a source here depicted as a $\chi^{(3)}$ OPO~\cite{Wu1986}. (b) Realization of a generalized mode-matching between the single-mode quantum states and the standard HD by an IME. For the single-mode case, this device is implemented through a a cavity resonant close to $\omega_0$. 
  (c) In the two-mode case both the LO of the standard HD and the quantum state are bi-color fields with carriers $\omega_1$ and $\omega_2$~\cite{Slusher1985}. (d) Our generalized mode-matching by an IME is  implemented through a coupled-cavity systems whose details are given in Section~\ref{sec:smooth dec}.\ref{subsec:Smooth_2_decomp}.}
  \label{fig:1_2_mode_detections}
\end{figure*}

In this work we address the  general case of multimode homodyne measurements. In this scenario
modematching takes on a new difficulty level due to the fundamental symmetries of HD,
which limit access to some spectral mode combinations resulting in mismatch-induced contamination~\cite{Barbosa2013}. This is particularly prevalent in quantum integrated photonics~\cite{Barbosa2018,PRXNice,Kogler2024}, as rich physics arises from the dissipative dynamics of optical parametric oscillators made of broadband, highly dispersive microresonators, and the ``black box'' nature of these systems requires sophisticated diagnosis and characterization tools. In particular, spectacular spectral squeezing features have been predicted, such as morphing supermodes~\cite{Gouzien2020} and hidden squeezing~\cite{Gouzien2023}. These exotic states are expected to be ubiquitously generated:
in microring resonators,
optomechanics, four-wave mixing (FWM) in atomic ensembles, polaritons in semiconductor microcavities, and quantum cascade lasers.
While spectral mode mismatch effects, at least in the absence of hidden squeezing, can be circumvented by making multiple different measurements on identical copies of the state, no solution has yet been found for making arbitrary single-shot measurements and new tools are needed.

One should note that pulse-shaping the HD LO doesn't solve the spectral mismatch problem because HD yields, by construction, the product of the respective temporal field modes of the LO and the quantum signal, whereas the product of the spectra is needed. Our approach therefore relies on implementing a convolution product in the time domain, 
using an ``interferometer with memory effect'' (IME), which can be mapped to a coupled-cavity array, as depicted in \fig{1_2_mode_detections}. 

Our approach extends HD capabilities to both frequency-dependent observables and their hidden correlations by using IMEs as an interface between HD and the quantum source. IMEs share a similar philosophy with, but represent a broader category of, quantum spectral engineering methods such as the production of frequency-dependent squeezing recently demonstrated in gravitational-wave interferometers~\cite{Miller2023,Kimble2001,McCuller2020}.

\section{Fundamental concepts}
\begin{figure*}[ht]
\centerline{\includegraphics[width=0.8\linewidth]{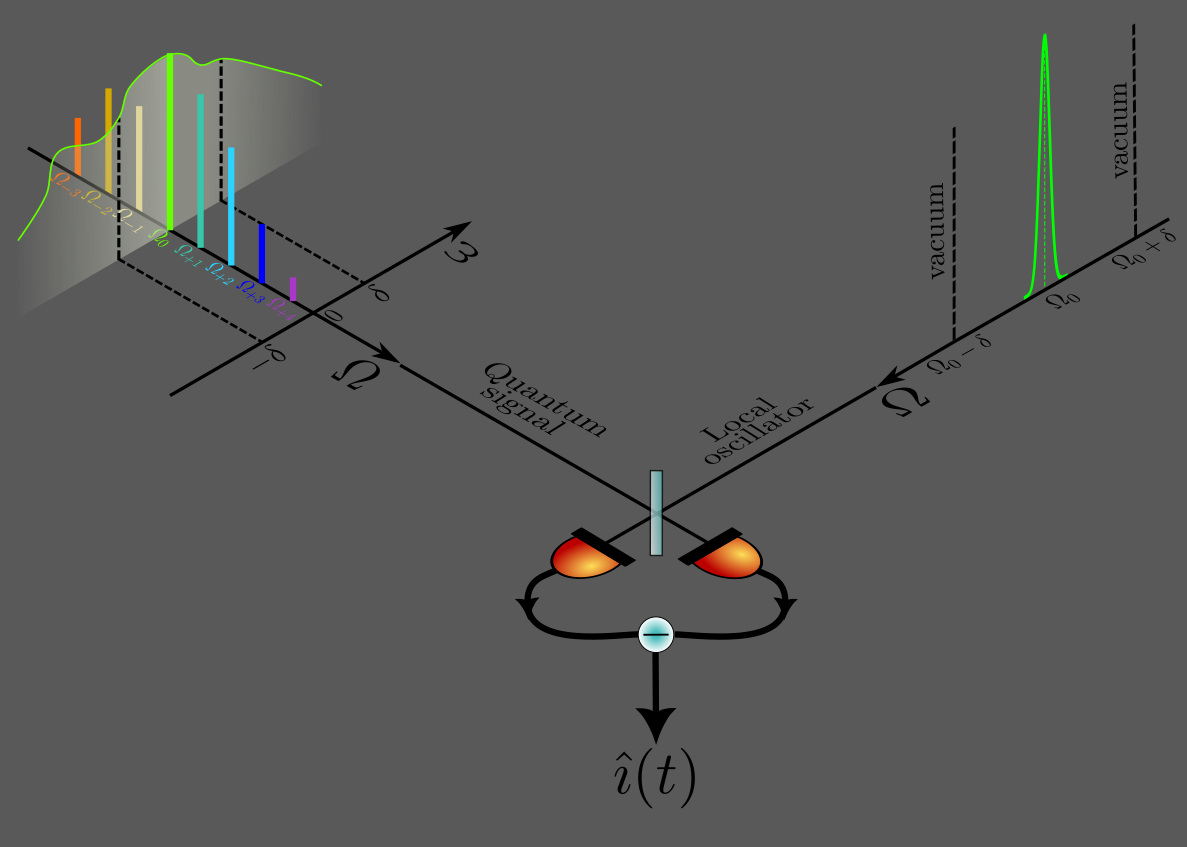}}
\caption{Principle of homodyne detection. Artist view of a situation where a morphing behavior is present. Note the spectral mismatch between the monochromatic LO and the broader-band quantum signal. In this configuration: (i) the LO with frequency-independent phase cannot project on the optimally squeezed quadratures that are frequency-dependent; (ii) {\it joint} measurements of quantum sidebands at $\omega_o\pm\delta$ can still be performed without vacuum noise contamination, but only for specific sideband symmetries, see text.}
\label{fig:HD}  
\vglue -.1in
\end{figure*}

\fig{HD} shows a basic example of spectral mismatch, very commonly encountered in squeezing experiments. The monochromatic LO at frequency $\Omega_o$ can sample the quantum signal at frequency $\Omega_o$ via a DC beat signal, i.e.\ an interference fringe, but other signal frequencies $\Omega_o\pm\delta$ will contribute to this same DC signal by interfering with the vacuum fields at these same frequencies, thereby leading to signal contamination by vacuum fluctuations. However, their contribution of such contamination to the whole DC fringe can generally be ignored when the LO amplitude is large compared to the quantum signal amplitude. The situation is notoriously less favorable when one considers frequency heterodyne measurements between the LO at $\Omega_o$ and a single quantum signal sideband at $\Omega_o+\delta$, resulting in a HD beat signal at electronic frequency $\delta$. Even for a large LO field, this signal is contaminated by the ``image'' vacuum sideband at $\Omega_o-\delta$. Frequency heterodyne measurements can still be gainfully used, however, in the case of a {\it two-mode} quantum signal at $\Omega_o\pm\delta$; in that case, the only vacuum contamination results from the contributions of the image vacuum sidebands~\cite{Yuen1980,Yuen1983} at $\Omega_o\pm2\delta$ which are negligible for LO amplitudes much larger than the quantum signal amplitudes. This approach can be extended to broadband, multimode measurements, e.g.~the squeezing spectrum, by scanning the electronic frequency $\delta$ at a given LO phase. However, that phase is the same for all values of $\delta$ and, since the universally accepted approach to this measurement is to assign a physical observable (quadrature) to each LO phase, this forces the measurement of the same quadrature for all values of $\delta$. This measurement is suboptimal since the optimally squeezed quadrature is frequency-dependent. 
While the recovery of the full information can be obtained through tomography by scanning the LO phase, this approach is not applicable to QIP and MBQC protocols, where the quantum resource must be accessed in one shot. Such tasks require frequency-dependent quadratures that are known as "morphing supermodes" in highly multimode situations~\cite{Gouzien2020}.

Another 
overarching issue with multimode HD is 
that its inherent symmetry only gives access to the variances and covariances of either one of the two EPR pairs ($x_s$,$y_a$) and ($x_a$,$y_s$), defined by joint field quadratures $q_{s,a}=q(\Omega_0+\delta)\pm q(\Omega_0-\delta)$, 
where $q$=$x$,$y$ respectively, $x$=($a$+$a^\dag$)/$\sqrt2$, and $y$=$i$($a^\dag$-$a$)/$\sqrt2$~\cite{Slusher1985}. This limitation means HD cannot provide measurements of both these pairs simultaneously~\cite{Barbosa2013,BarbosaPRA2013}. As a consequence, HD fails to access quantum cross-correlations between, say, $x_s$ and $x_a$, leading to a phenomenon termed ``hidden squeezing''~\cite{Gouzien2023}.

\section{Quadratic Hamiltonian cavity quantum optics}
Any multimode optical field can be decomposed over a orthonormal family of modes
$\{\bm{u}_n(\bm{r},t)\}$, involving different degrees of freedom (spatial, time, frequency, polarization), each associated to boson operators $\hat{a}_m ,\hat{a}^\dagger_m $ satisfying the commutation relations $[ \hat{a}_m, \hat{a}^\dagger_n]=\delta_{m,n}$ and $[\hat{a}_m, \hat{a}_n]=0$ . With this notation, the positive frequency part of the electric field reads as~\cite{FabreReview}
\begin{equation}
\hat{\bm{E}}^{(+)}(\bm{r},t)=\sum_m \mathcal{E}_m \hat{a}_m \bm{u}_m(\bm{r},t),
\end{equation}
where bold letters represent vector quantities and $\mathcal{E}_m$ is the single photon electric field.
Here, we will focus on the case of longitudinal modes of an optical cavity, all sharing the same spatial transverse mode and linear polarization: 
\begin{equation}
\bm{u}_m(\bm{r},t)\rightarrow\bm{\epsilon}\exp{i(k_m z-\Omega_m t)}.
\end{equation}
This covers the very broad category of multi-color quantum states and quantum frequency combs out of optical cavities or single-mode waveguides~\cite{Pfister2019}. 
Nevertheless, the principles of generalized mode-matching discussed here could be easily extended to other kinds of modes.

In Heisenberg picture, the nonlinear dissipative evolution of a system of $N$ bosonic modes is described by a system of coupled quantum Langevin equations. A standard linearization procedure around a stable stationary solution allows reducing such evolution to the effective Hamiltonian
\begin{equation}\label{ham}
\hat{H}=\hbar\sum_{m,n} G_{m,n} \hat{a}_m^{\dag}\hat{a}_n +\frac{\hbar}{2}\sum_{m,n}\left[F_{m,n} \hat{a}_m^{\dag}\hat{a}_n^{\dag} + \hc{}\right]
\end{equation}
which represents the most general time-independent quadratic Hamiltonian. The $N\times N$ matrices $G$ and $F$ incorporate the details of the linearized dynamics of the considered physical system and verify, respectively,  $F=F^\transp$, and $G=G^{\dagger}$\footnote{We are using the following notation: ${[\cdot]}^\transp$ for the transpose, ${[\cdot]}^*$ 
for the complex conjugate and ${[\cdot]}^\dagger$ for the Hermitian transpose.}.
In practical situations, the matrix $F$ has the nature of pairs production, as those arising from spontaneous parametric down-conversion in $\chi^{(2)}$ or $\chi^{(3)}$ interactions
with undepleted pumps~\cite{Patera2010, Chembo2016}. The very general shape of the matrix $G$ takes into account various mode-hopping processes whose details depend on the nature of the boson modes considered.
In the case of cavity longitudinal modes, $G$ includes frequency conversion processes~\cite{Christ2011}, 
$\chi^{(3)}$-induced self- and cross-phase modulation~\cite{Chembo2016}, mode detunings from perfect resonance and linear dispersion effects. 
~\eqref{ham} generates the set of linear quantum Langevin equations
\begin{align}
\frac{\mathrm{d}\hat{\bm{R}}(t)}{\mathrm{d}t}&=
(-\Gamma+\mathcal{M})\hat{\bm{R}}(t)+
\sqrt{2\Gamma}\,\hat{\bm{R}}_{\mathrm{in}}(t)
\label{langevin quad},
\end{align}
where $\hat{\bm{R}}(t)=(\hat{\bm{x}}(t)|\hat{\bm{y}}(t))^\transp$, $\hat{\bm{x}}(t)=(\hat{x}_{1},\ldots,\hat{x}_{N})^\transp$ and $\hat{\bm{y}}(t)=(\hat{y}_{1},\ldots,\hat{y}_{N})^\transp$ 
are the column vectors of the amplitude and phase quadratures of the intracavity modes, $\hat{x}_m=(1/\sqrt{2})(\hat{a}_m^\dag+\hat{a}_m)$ and $\hat{y}_m=(\mathrm{i}/\sqrt{2})(\hat{a}_m^\dag-\hat{a}_m)$. The diagonal matrix $\Gamma$ describes mode-dependent cavity losses and the mode coupling matrix $\mathcal{M}$ is expressed as
\begin{equation}
\mathcal{M}=
\left(
\begin{array}{c|c}
\mathrm{Im}\left[G+F\right] & \mathrm{Re}\left[G-F\right]
\\
\hline
-\mathrm{Re}\left[G+F\right] & -\mathrm{Im}\left[G+F\right]^\transp
\end{array}
\right)
\label{eMMe}
\end{equation}
where $\Re[G-F]$ and $\Re[G+F]$ symmetric matrices.

The quadratures $\hat{\bm{R}}_{\mathrm{out}}$ of the multimode state at the device output can be conveniently obtained as functions of input modes' quadratures, thanks to standard input-ouput relations $\hat{\bm{R}}_{\mathrm{in}}+\hat{\bm{R}}_{\mathrm{out}}=\sqrt{2\Gamma}\,\hat{\bm{R}}$~\cite{GardinerZoller}. In the Fourier space, they are given by
\begin{align}\label{pene}
\hat{\bm{R}}_{\mathrm{out}}(\omega)&=
S(\omega)\hat{\bm{R}}_{\bm{in}}(\omega),
\end{align}
where $S(\omega)$ is the system's transfer function. It is important to note here that we work in the interaction picture with fields in the slowly varying amplitude approximation. Therefore, $\omega$ can be taken to rigorously coincide with the sideband frequency offset from the optical carriers at $\Omega_m$. $S(\omega)$ is a complex matrix-valued function of the continuous parameter $\omega$ given by the expression
\begin{equation}
S(\omega)=\sqrt{2\Gamma}\left(\mathrm{i}\omega\mathbb{I}+\Gamma-\mathcal{M}\right)^{-1}\sqrt{2\Gamma}-\mathbb{I}
\label{S}
\end{equation}
with $\mathbb{I}$ the $2N\times 2N$ identity matrix. The input spectral quadrature operators satisfy the commutation rule $\left[\hat{\bm{R}}_{\mathrm{in}}(\omega),\hat{\bm{R}}_{\mathrm{in}}^{\transp}(\omega')\right]=i\mathbb{J}\delta(\omega+\omega')$, 
where $\mathbb{J}=\begin{pmatrix} 0 & I \\  -I & 0\end{pmatrix}$ is the $N$-mode symplectic form and $I$ is the $N\times N$ identity matrix. The output quadratures $\hat{\bm{R}}_{\mathrm{out}}(\omega)$ are the Fourier transform of \textit{bona fide} boson quadrature operators in time domain, since $S(\omega)$ has the two following properties~\cite{Gouzien2020}:
\begin{property}
$S$ is conjugate symmetric: $S(-\omega)=S^*(\omega)$. This assures the reality of $S$ in time domain.
\end{property}
\begin{property} 
$S$ is ``$\omega$-symplectic'' which, for the sake of succinctness, stands for a smooth (analytic) matrix-valued function that is conjugate-symplectic\footnote{The conjugate-symplectic group Sp$^*(2N,\mathbb{C})$ is defined as the set of $2N\times 2N$ complex matrices such that $S\mathbb{J} S^\dagger$, with $\mathbb{J}$ the symplectic form.} for all $\omega$. 
This can be proved from the property of $\mathcal{M}$ of being an Hamiltonian matrix $[{(\mathbb{J}\,\mathcal{M})}^{\transp}=\mathbb{J}\,\mathcal{M}]$
and of $\Gamma$ of being a skew-Hamiltonian $[{(\mathbb{J}\,\Gamma)}^{\transp}=-\mathbb{J}\,\Gamma]$;
\end{property}
\subsection{Covariance matrix of the multimode state}
The transfer function $S(\omega)$ is in general non diagonal, leading to a coupling between the quadratures of different output modes. This reflects the presence, at the output of the device, of continuous variable entanglement among field optical modes at frequencies $\Omega_m$. In other words, when measuring the quadrature $\hat{\bm{R}}_{\mathrm{out}}$, a correlation appears among noise fluctuations of the different optical modes. In time domain, a quantification of such correlation is provided by the covariance matrix
$\sigma_{\mathrm{out}}(t)=\frac{1}{2}\langle \hat{\bm{R}}_{\mathrm{out}}(0)\hat{\bm{R}}_{\mathrm{out}}^{\transp}(t)
+{(\hat{\bm{R}}_{\mathrm{out}}(t)\hat{\bm{R}}^{\transp}_{\mathrm{out}}(0))}^{\transp}\rangle$ that, by assuming input vacuum states $\hat{\bm{R}}_{\mathrm{in}}$,
 fully characterizes the stationary multimode Gaussian state at the micro-resonator output~\cite{Ferraro_book}. 
The transfer function $S(\omega)$ allows calculating in a direct way, in Fourier domain, the spectral covariance matrix as:
\begin{equation}\label{spectralcov}
\sigma_{\mathrm{out}}(\omega)
= \frac{S(\omega)S^{\transp}(-\omega)}{2\sqrt{2\pi}}.
\end{equation}
This is, in general, a non-real Hermitian matrix and, from property 1, conjugate symmetric, \textit{i.e.} $\sigma(-\omega)=\sigma^{*}(\omega)$; as a consequence the real part of the spectral covariance matrix is symmetric with respect to the exchange $\omega\leftrightarrow -\omega$, $\mathrm{Re}[\sigma(-\omega)]=\mathrm{Re}[\sigma(\omega)]$, 
and its imaginary part is anti-symmetric, $\mathrm{Im}[\sigma(-\omega)]=-\mathrm{Im}[\sigma(\omega)]$. 
\bigskip

The complex nature of the spectral covariance matrix is generally ignored in the literature \cite{Dauria2010, Laurat2005,Chembo2016,Guidry2023} and only the real part of~\eqref{spectralcov} is typically considered. This assumption leads, however, to disregarding the presence of correlations among noise components that emerge from the properties of complex covariance matrices $\sigma_{\mathrm{out}}(\omega)$. From a physical point of view, the full characterization of the quantum state should involve measurement of both quantum noise components at frequencies $\omega$ and $-\omega$. 
However, in some situations~\cite{Fabre1989} the qumodes  at $\omega$ and $-\omega$ are essentially in the same quantum state and only the subspace $\omega$ (or the subspace $-\omega$) can be considered: such a state symmetry with respect to $\omega\leftrightarrow -\omega$ makes the assumption of a real $\sigma_{\mathrm{out}}(\omega)$ correct and justifies the use of standard homodyne detector to fully retrieve the state quantum properties.

A complex $\sigma_{\mathrm{out}}(\omega)$ implies, instead, a state that is not symmetric with respect to qumodes at $\omega$ and $-\omega$. For those states, considering only qumodes at $\omega$ means neglecting quantum correlations between them and those at $-\omega$, that are encoded in the imaginary part of the spectral covariance matrix. As it will be detailed in the following, their measurement is not accessible to a standard homodyne detector and rather requires novel strategies. These extra correlation are essential for fully characterizing the corresponding quantum state and could also be exploited for enhancing the performances of QIP and MBQC protocols.

\subsection{Morphing supermodes}

In experiments, the caracterization of the covariance matrix of a multimode state requires the simultaneous measure of the noise features of the different optical modes and then retrieving their correlation. This strategy, although quite current in two mode scenarios\,~\cite{Laurat2005}, is not viable when the number of correlated modes scales up. In many practical situations, a preferred way to proceed is to introduce a set of normal modes (a.k.a. supermodes) that decouple the system dissipative dynamics and map the CV multimode entangled state 
into a collection of (anti-)squeezed states that, since they are statistically uncorrelated, can be measured independently of each other~\cite{Patera2010, Roslund2014}. As demonstrated in Ref.~\cite{Gouzien2020}, in the general case of a quadratic Hamiltonian, the modes containing the optimal (anti-)squeezing are ``morphing''. 
Morphing supermodes are a generalisation of standard (static) supermodes and are expressed as linear combinations of the initial system optical modes with \textit{frequency-dependent} coefficients that vary with $\omega$. A morphing behaviour is quite common in nonlinear optical cavities. It appears, for instance, in single-mode degenerate optical parametric oscillators with signal or pump detuning~\cite{Fabre1989,Fabre1990,Solimeno2002} and in optomechanical cavities~\cite{Fabre1994,Mancini1994,Junker2022}. At their output, optimal squeezing is found in a frequency-dependent quadrature that varies at scales of the cavity bandwidth. Surprisingly, it was only recently that it has been unveiled in highly multimode systems such as silicon-based microresonators~\cite{Gouzien2023}. It also appears in the description of non-Gaussian states generated 
in ultra-fast second-order nonlinear photonics
beyond the pump depletion approximation~\cite{Yanagimoto2022,Jankowski2024}.

The morphing supermodes can be formally obtained
through the ``analytic Bloch-Messiah decomposition'' (ABMD) of the transfer function~\cite{Gouzien2020}
\begin{align}\label{ABMD}
S(\omega)=U(\omega)D(\omega)V^{\dag}(\omega)
\end{align}
where $U(\omega)$ and $V(\omega)$ are unitary and $\omega$-symplectic matrix-valued functions that characterize the structure of the morphing supermodes while the diagonal matrix-valued function   
$D(\omega)= \diag\{d_1(\omega),\ldots,d_N(\omega)|\,d_1^{-1}(\omega),\ldots,d_N^{-1}(\omega)\}$ (with $d_i(\omega)\ge1$ for $i=\{1,\ldots,N\}$,
for all $\omega$) contains their (optimal) anti-squeezing $d_i(\omega)$ and squeezing $d_i^ {-1}(\omega)$ levels. All of these functions inherit Property 1 of the transfer matrix, making them conjugate symmetric. In general, ABMD differs from a simple point-by-point Bloch-Messiah decomposition for each $\omega$, as it ensures the smoothness of supermodes that allows to define the continuously evolving physical observables
\begin{equation}\label{morsup}
\hat{\bm{R}}^{(s)}_{\mathrm{out}}(\omega) = \bm{U}^\dagger(\omega)\hat{\bm{R}}_{\mathrm{out}}(\omega)
\end{equation}
that represent the optimally (anti-)squeezed quadrature operators. Their noise level is given by $d_i(\omega)$ for the anti-squeezed supermode $i$ and by $d_i^ {-1}(\omega)$ for squeezed supermode $N+i$.
Since $U(\omega)$ is $\omega$-symplectic and conjugate symmetric, the new quadratures have a symmetric real part and anti-symmetric immaginary part
so to correspond to Hermitian quadratures in time domain.

\subsection{Hidden squeezing}
As for the covariance matrix, morphing supermodes are complex-valued for of quantum states exhibiting asymmetric behaviour with respect to the exchange $\omega\leftrightarrow -\omega$. This results in quantum correlations that cannot be detected by standard HD, a phenomenon that has been named ``hidden squeezing"~\cite{Gouzien2023}.
The concept of hidden squeezing was first explored and explained by Barbosa \textit{et al.} (see for example~\cite{Barbosa2013,BarbosaPRA2013}). In their works, the authors pointed out that the squeezing of a single beam of light actually involves two-mode squeezing correlations between symmetric noise sidebands at $\omega$ and $-\omega$. An effective single-mode description can be applied when the noise spectral power of these symmetrical components are identical, while in general, a two-mode description becomes necessary: in such cases, homodyne detection lacks the necessary degrees of freedom to detect correlations between the symmetrical components and, as a consequence, to reconstruct the complete information~\cite{Barbosa2013}. The solution proposed by Barbosa and coworkers showed that a "resonator detection'' (RD) successfully detects correlation between noise spectral components at $\omega$ and $-\omega$ ~\cite{Barbosa2013, BarbosaPRA2013}. However, such architecture does not easily scale with the number of entangled modes nor can cope with the frequency-dependent behaviour of the morphing supermodes.

The characterization of the covariance matrix, either in the original basis of the longitudinal modes or in the supermodes basis, require the detection of mode quadratures, \textit{i.e.} of the field continuous variables~\cite{Dauria2010, Laurat2005, Roslund2014}. In experiments, quadratures are typically accessed via HD by making the fields under scrutiny beat with a bright reference beam called ``local oscillator" (LO)~\cite{Bachor2019}. The optical beat signal, after photodetection, gives a photocurrent that encodes the quadrature mean value and quantum noise~\cite{Bachor2019}. For quantum states generated in the longitudinal modes of a cavity, the LO takes the following form
\begin{align}
E_{\mathrm{LO}}(t)&=\mathrm{i}\sum_{m}\alpha_m\,\mathrm{e}^{-\mathrm{i}\Omega_m t}+\mathrm{H.c.}\label{LO}
\end{align}
where each $\alpha_m$ is the complex amplitude of the LO components in the optical mode at frequency $\Omega_m$. Correspondingly, the photocurrent operator can be written as
\begin{align}
\hat{\imath}(t)&\propto
\sum_m
\Big(
\mathrm{Re}[\alpha_m]\, \hat{x}_{\mathrm{out},m}(t)+ \mathrm{Im}[\alpha_m]\, \hat{y}_{\mathrm{out},m}(t)
\Big)=
\nonumber\\
&=\bm{Q}^\transp \hat{\bm{R}}_{\mathrm{out}}(t).\label{photocurrent}
\end{align}
By defining $\bm{\alpha}=\left(\alpha_{1},\ldots,\alpha_{N}\right)^\transp$, in this expression, $\bm{Q}=\left(\mathrm{Re}[\bm{\alpha}]\,|\,\mathrm{Im}[\bm{\alpha}]\right)^\transp$ is the normalised column vector of LO coefficients in the quadrature representation. Although LOs with arbitrary spectral amplitudes $\bm{\alpha}$ are accessible to experiments~\cite{Roslund2014}, the amplitudes and phases of their spectral components always combine to give real quadratures, \textit{i.e} a real $\bm{Q}$: this guarantees that, in time domain, the photocurrent operator~\eqref{photocurrent} is Hermitian. 
The projective measurement described by~\eqref{photocurrent} gives a combination of the quadratures of optical fields at frequencies $\Omega_m$ with coefficients given by $\bm{Q}$: by shaping the LO it is thus possible to measure the quadratures of individual longitudinal modes or of their linear combinations (for example in the basis of the supermodes). For frequency homodyning~\cite{Bachor2019}, the Fourier transform of the photocurrent returns the noise spectrum of the chosen quadrature:
\begin{equation}\label{noisespectrum}
\Sigma_{\bm{Q}}(\omega)
= \bm{Q}^\transp\sigma_{\mathrm{out}}(\omega)\bm{Q}.
\end{equation}
As it should be for any measurable quantity, this expression always returns a real noise spectrum $\Sigma_{\bm{Q}}(\omega)$ (as shown in Appendix~\ref{ap:NSP}), even when the covariance matrix $\sigma_{\mathrm{out}}(\omega)$ exhibits an imaginary part. 

When dealing with quantum states that have complex spectral covariance matrix, the limitations of standard HD can be illustrated with the very simple case of a single mode state described by its amplitude and phase quadratures $(\hat{x},\hat{y})^\transp$,
\begin{align}
\sigma_{\mathrm{out}}(\omega)&=
\left(
\begin{array}{cc}
\alpha(\omega) & \gamma(\omega)+\mathrm{i}\delta(\omega)
\\
\gamma(\omega)-\mathrm{i}\delta(\omega) & \beta(\omega)
\end{array}
\right)
\end{align}
where $\alpha(\omega)$, $\beta(\omega)$, $\gamma(\omega)$ and $\delta(\omega)$ are real functions of $\omega$ as required to guarantee the Hermicity of $\sigma_{\mathrm{out}}(\omega)$. The corresponding noise spectral power~\eqref{noisespectrum}, as detected from HD, corresponds to
\begin{align}
\Sigma_{\bm{Q}}(\omega)&=
\cos^2\theta\,\alpha(\omega)
+
\sin^2\theta\,\beta(\omega)
+
2\cos\theta\sin\theta\,\gamma(\omega)
\end{align}
for a given LO, $\bm{Q}=(\cos\theta|\sin\theta)^\transp$. Whatever the choice of $\theta$ (\textit{i.e.} of the selected quadrature), $\Sigma_{\bm{Q}}(\omega)$ clearly does not encode any information about the imaginary part of the spectral covariance matrix $\delta(\omega)$ that, therefore, carries quantum correlations that are hidden to HD.

In the case of an arbitrary (large) number of modes, the spectral covariance matrix is easier to characterize in terms of the optimal (anti-)squeezing levels associated with individual supermodes. In this basis, the limitations of the HD in dealing with Gaussian states with complex $\sigma(\omega)$ and with morphing behaviour can be formalized as a problem of mode-matching: the occurrence of sub-optimal detection of squeezing originates from a LO not perfectly projecting on the targeted supermode.
By replacing~\eqref{ABMD} in~\eqref{noisespectrum}, the detected noise spectral power takes the following form
\begin{equation}\label{noisespectrum2}
\Sigma_{\bm{Q}}(\omega)
=\frac{1}{2\sqrt{2\pi}}
\, 
[\bm{Q}^\transp U(\omega)]
\,D^2(\omega)
\,
[U^{\dagger}(\omega)\bm{Q}].
\end{equation}
This expression shows that to measure the optimal degree of squeezing, \textit{i.e.} 
$\Sigma_{\bm{Q}}(\omega)\sim d_i^{-2}(\omega)$, the LO should perfectly match the $(N+i)$-th supermode. This corresponds to achieve a perfect projection on the supermode identified by the $(N+i)$-th column of $U(\omega)$, $\bm{U}_{N+i}(\omega)$. Such a projection formally corresponds to shape the LO column vector $\bm{Q}$ so that the scalar product $\bm{Q}^\transp \bm{U}_{N+i}(\omega)=1$.
However, this mode-matching condition cannot be satisfied as the LO $\bm{Q}$ is constant and real while, in general, $U$ depends on the frequency $\omega$ and it is complex. In the special case of a real $U(\omega)$, a LO with a non-morphing profile (corresponding a frequency independent $\bm{Q}$), can retrieve the optimal squeezing only at the frequency $\bar{\omega}$ 
for which $\bm{Q}$ and $\bm{U}_{N+i}(\bar{\omega})$ are matched. In the general case of a complex $U(\omega)$, no mode-matching is possible and the HD measure is suboptimal for \textit{any} values of $\omega$ ~\cite{Gouzien2023}, thus leaving a part of the quantum correlations hidden. 

Limited access to squeezing as well as the impossibility of following the morphing behaviour are extremely detrimental in 
practical situations. Many quantum optical protocols require homodyne to be performed in time domain: in these cases, the detection signal is acquired as a function of the time over a certain time slot. In Fourier domain, such a signal simultaneously carries contributions from spectral components at different $\omega$s~\cite{Lvovsky2013}. Working with a non-morphing LO means projecting the detected field onto the \textit{same} linear combination of longitudinal modes whatever $\omega$. Even if such a linear combination would match a squeezed supermode at $\bar{\omega}$, this is a priori not the case for all other frequency components. As a consequence, time-homodyne signal mixes noise levels associated to the optimal squeezing at $\bar{\omega}$, with the sub-optimal or even anti-squeezed noise contributions associated to all other Fourier components. As understandable, the presence of hidden squeezing further worsens the situation. The inability to work with standard homodyne detection affects a wide range of situations where time-domain homodyning is essential, including heralded state preparations~\cite{Lvovsky2013, melalkia_plug-and-play_2022, RaMultiNG2020} or MBCQ~\cite{Menicucci2014ft}, making it urgent to identify a different, better adapted detection strategy.

\section{Achieving generalized quantum mode-matching}\label{sec:generalized mode-matching}
\begin{figure}[h!]
  \centering
  \includegraphics[width=1\linewidth]{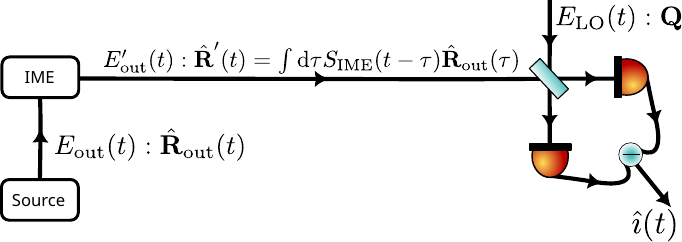}
  \caption{This scheme outlines the generalized mode-matching through an interferometer with memory. In the first stage, the physical system generates the CV quantum resource, with its output denoted as $\hat{E}{\mathrm{out}}(t)$ and associated with the vector of quadratures $\hat{\bm{R}}{\mathrm{out}}(t)$. This output then undergoes the second stage, which is an IME characterized by a transfer function $S_{\mathrm{IME}}(\omega)$. Subsequently, the IME output, characterized by the vector of quadratures $\hat{\bm{R}}'_{\mathrm{out}}(t)$, is directed into a traditional HD scheme with a LO described by the vector $\bm{Q}$.}
  \label{fig:IME}
\end{figure}

In order to project onto the $i$-th morphing supermode, not only one should be able to produce a LO associated 
to a complex and frequency dependent $\tilde{\bm{Q}}(\omega)$
\begin{equation}
\bm{Q}\rightarrow\tilde{\bm{Q}}(\omega),
\end{equation}
but also to implement, in frequency domain, the multiplication between this LO and the targeted field, so as to induce a linear combination giving
the quadrature, $\hat{R}^{(s)}_{\mathrm{out},i}(\omega)$, of the $i$-th morphing supermode
\begin{equation}\label{Qproduct}
\hat{R}^{(s)}_{\mathrm{out},i}(\omega)=\tilde{\bm{Q}}^\dagger(\omega) \hat{\bm{R}}_{\mathrm{out}}(\omega)
\end{equation}
when $\tilde{\bm{Q}}(\omega)$ is shaped as $\bm{U}_{i}(\omega)$. A simple shaping of the LO in order to produce a time dependent profile with Fourier transform $\tilde{\bm{Q}}(\omega)$ would not be sufficient. Indeed, in a standard HD, the mixing of the LO and the signal field at a beamsplitter rather leads to a multiplication in time domain. Therefore this approach, know as "synodyne detection'' and proposed by Buchmann \textit{et al.} in~\cite{Buchmann2016} for the HD of single mode ponderomotive squeezing, is able to mode-match complex supermodes but not their morphing behaviour.

To obtain, in frequency domain, a transformation such as~\eqref{Qproduct}, we must look for a physical system that implements, in time domain, the following transformation of the signal field
\begin{equation}\label{Qconv}
\hat{R}^{(s)}_{\mathrm{out},i}(t)=
\int\mathrm{d}\tau\,
\tilde{\bm{Q}}^\transp(t-\tau)\hat{\bm{R}}_{\mathrm{out}}(\tau).
\end{equation}
In other words, we need a physical system that is linear and possesses both memory, as optical cavities, and the capability to mix the input quadratures, as interferometers. 
These two characteristics define a device we refer to as an "interferometer with a memory effect'' (IME). 

The transformation~\eqref{Qproduct} or, equivalently,~\eqref{Qconv} can be physically implemented by cascading, at the output of the source of quantum light, an IME and a standard HD, 
as depicted in \fig{IME}.
The IME transforms the quadratures, $\hat{\bm{R}}_{\mathrm{out}}(\omega)$, of the output of the multimode source according to the relation:
\begin{align}
\hat{\bm{R}}'(\omega)=S_{\mathrm{IME}}(\omega) \hat{\bm{R}}_{\mathrm{out}}(\omega),
\end{align}
where $S_{\mathrm{IME}}(\omega)$ is the IME transfer function. The HD photocurrent operator, thus, becomes:
\begin{align}\label{photocurrent2}
\hat{\imath}(\omega)&\propto
\tilde{\bm{Q}}^\dagger(\omega)\hat{\bm{R}}_{\mathrm{out}}(t),
\end{align}
with a generalized LO given by 
\begin{equation}\label{gen LO}
\tilde{\bm{Q}}^\dagger(\omega) = \bm{Q}^\transp S_{\mathrm{IME}}(\omega).
\end{equation}
The combination of a tailored LO, with real and non-morphing $\bm{Q}$, and of a suitably engineered $S_{\mathrm{IME}}(\omega)$ allows $\tilde{\bm{Q}}(\omega)$ to match $\bm{U}_{N+i}(\omega)$ as required to fully detect the squeezing of the $(N+i)$-th morphing supermode.
\begin{figure*}[ht!]
  \centering
  \includegraphics[width=0.8\linewidth]{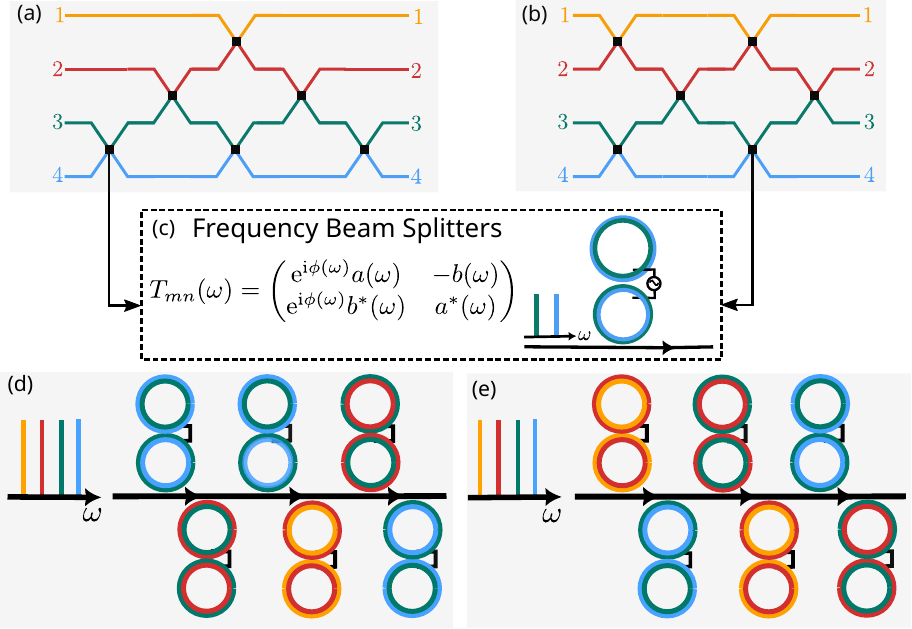}
  \caption{Smooth two-mode decomposition of a 4-mode passive transformation with (a) triangular mesh and (b) rectangular mesh. Two-mode unitary (c) and its physical implementation via a two-mode coupled cavity. Implementation in an integrated platform of microresonators of respectively the (d) triangular and (e) rectangular decomposition.}
  \label{fig:two_mode_meshes}
\end{figure*}

Since the IME stage must not change the amount of squeezing (anti-squeezing), this system should implement a passive transformation ($F=0$ in~\eqref{ham} and~\eqref{eMMe}). Thus, the corresponding transfer function that solves the quantum Langevin equations~\eqref{S}, in Fourier domain, reads as:
\begin{equation}
S_{\mathrm{IME}}(\omega)=\sqrt{2\Gamma_{\mathrm{IME}}}\left(\mathrm{i}\omega\mathbb{I}+\Gamma_{\mathrm{IME}}-\mathcal{M}_{\mathrm{IME}}\right)^{-1}\sqrt{2\Gamma_{\mathrm{IME}}}-\mathbb{I}
\label{S_IME}
\end{equation}
with mode-dependent losses matrix $\Gamma_{\mathrm{IME}}$ and the coupling matrix given by
\begin{equation}
\mathcal{M}_{\mathrm{IME}}=
\left(
\begin{array}{c|c}
\mathrm{Im}\left[G_{\mathrm{IME}}\right] & \mathrm{Re}\left[G_{\mathrm{IME}}\right]
\\
\hline
-\mathrm{Re}\left[G_{\mathrm{IME}}\right] & -\mathrm{Im}\left[G_{\mathrm{IME}}\right]^\transp
\end{array}
\right).
\label{eMMe_IME}
\end{equation}
At the output of the IME the state is characterized by the spectral covariance matrix
\begin{align}
\sigma'(\omega)
&=
S_{\mathrm{IME}}(\omega)\sigma_{\mathrm{out}}(\omega)S_{\mathrm{IME}}^{\transp}(-\omega),
\label{spectralcov2}
\end{align}
and the noise spectral power obtained by the HD after the IME is
\begin{equation}\label{noisespectrum3}
\Sigma_{\bm{Q}}(\omega)
=\frac{1}{2\sqrt{2\pi}}
\, 
[\tilde{\bm{Q}}^\dagger(\omega) U(\omega)]
\,D^2(\omega)
\,
[U^{\dagger}(\omega)\tilde{\bm{Q}}(\omega)]
\end{equation}
where $\tilde{\bm{Q}}(\omega)$ is given by expression~\eqref{gen LO}.

In particular, mode-matching can even be realized simultaneously for all the frequency-dependent supermodes through a unique IME configuration by requiring that
\begin{equation}
\frac{d}{d\omega}\Big[S_{\mathrm{IME}}(\omega) U(\omega)\Big]=0.
\label{stationary MM}
\end{equation}
As a special case, this expression admits the equality $S_{\mathrm{IME}}(\omega) U(\omega)=I$, but other solutions are possible.
However, it is clear that trying to find a IME that satisfies~\eqref{stationary MM} represent a much harder optimization problem.

As a concluding remark, it is pertinent to the note that, for single mode system, an IME consists in an empty cavity with detuning $G_{\mathrm{IME}}=\Delta$. In this case the scheme corresponds to 
the resonator detection (RD) as in~\cite{Barbosa2013}. The RD is the simplest form of IME, our results thus show that RD can also project onto morphing supermodes, a property that was not recognized so far. 
Furthermore, IME implements frequency-dependent multimode linear transformation that can contain also frequency-conversion terms. 
These could be physically implemented by electro-optic modulators or nonlinear optical coupling, as it will be described in the following section.
As a consequence, with respect to RD, IMEs represent a broader and more general class of devices for the manipulation and characteriztion of quantum states.

\section{Implementing the interferometer with memory effect}\label{sec:smooth dec}
The IME is a passive, linear system that is able to implement an arbitrary transformation described by an frequency-dependent unitary $S_{\mathrm{IME}}(\omega)$. To identify the optical system implementing the IME, we introduce an original technique to decompose any $S_{\mathrm{IME}}(\omega)$ in terms of networks of basic photonic components arranged in specific architectures: changing the components' parameters allows realizing different $S_{\mathrm{IME}}(\omega)$. To detect a target squeezed morphing supermode, such parameters are numerically optimized to make $\tilde{\bm{Q}}^\dagger(\omega)$ match $\bm{U}_{N+i}(\omega)$ in~\eqref{gen LO} .
Our method extends ideas introduced in the context of discrete spatial modes~\cite{Li2013} in order to find decompositions that are smooth with respect to $\omega$. 

For practical purposes, we approach this problem in the representation $\hat{\bm{\xi}}=(\bm{a}|\bm{a}^\dagger)^\transp$ of the complex field amplitudes, 
where $\hat{\bm{a}}=(\hat{a}_1,\cdots,\hat{a}_{\mathrm{N}})^\transp$. In this case, the transfer function of the IME is given by 
$\mathbb{U}_{\mathrm{IME}}(\omega) = L^{\dagger} S_{\mathrm{IME}}(\omega) L$, with $L=\frac{1}{\sqrt{2}}\begin{pmatrix} I & I \\ -i I & i I \end{pmatrix}$, $I$ being the identity matrix, and it assumes the form
\begin{equation}
\mathbb{U}_{\mathrm{IME}}(\omega)=
\left(
\begin{array}{c|c}
U_{\mathrm{IME}}(\omega) & 0
\\
\hline
0 & U^*_{\mathrm{IME}}(-\omega)
\end{array}
\right),
\end{equation}
where $U_{\mathrm{IME}}(\omega)$ is a $N\times N$ smooth unitary matrix-valued function of $\omega$.
In the following subsections we will discuss its frequency-dependent decomposition into elementary components. 

\begin{figure*}[t!]
  \centering
  \includegraphics[width=\linewidth]{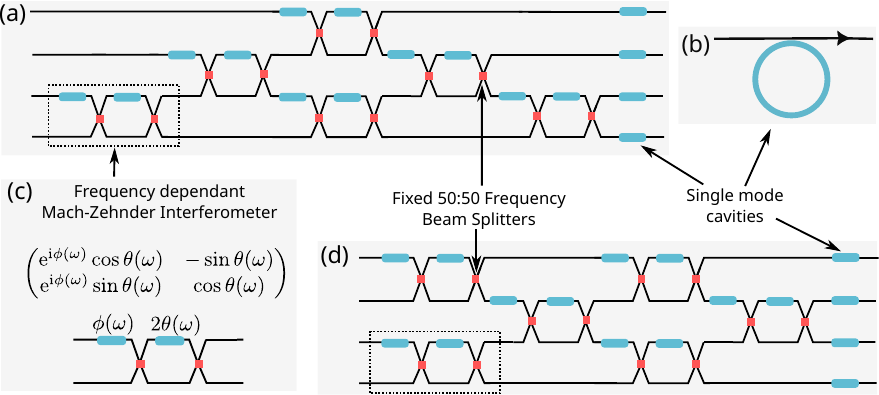}
  \caption{(a) Smooth single-mode decomposition of a 4-modes passive transformation with triangular mesh. (b) Single mode cavity. (c) Frequency dependent Mach-Zehnder interferometer. (d) Smooth single-mode decomposition with rectangular mesh.}
  \label{fig:meshes}
\end{figure*}
\subsection{Smooth two-mode decomposition} \label{subsec:Smooth_2_decomp}
A first possible decomposition of the matrix $U_{\mathrm{IME}}(\omega)$ can be done in terms of elementary bricks that corresponds to 2-modes passive transformations. 
Specifically, we show that frequency-dependent unitary transformations can be systematically synthesized from basic components through the following proposition:
\begin{proposition}
Any $N\times N$ frequency-dependent unitary can be decomposed into a mesh of at most $N(N-1)/2$ 
two-mode frequency-dependent unitaries $T_{mn}(\omega)$  (\fig{two_mode_meshes}c).
\end{proposition}
Therefore,
\begin{equation}
U_{\mathrm{IME}}(\omega) = \prod_{(m,n)}T_{mn} (\omega),
\label{smooth decomp}   
\end{equation}
where the $T_{mn}(\omega)$ are the most general $N\times N$ unitary transformations that leave uncoupled all the modes with the exception of the two modes labelled with indices $m,n \in\{1,2, \cdots, N\}$. The explicit and full expression of these matrices, known in the literature as ``two-level unitaries", is given in Appendix~\ref{ap:Smooth_decomp}. Here, for the sake of brevity and with a (minor) abuse of notation, we will use the following $2\times 2$ form: 
\begin{equation}
T_{mn}(\omega)=
\begin{pmatrix} 
\mathrm{e}^{\mathrm{i} \phi(\omega)}a(\omega)& -b(\omega) \\ 
\mathrm{e}^{\mathrm{i} \phi(\omega)}b^*(\omega)& a^*(\omega)
\end{pmatrix}.\label{Tmn}
\end{equation}
In the the quadratures representation, matrices $T_{mn}(\omega)$ are given by eqs.~\eqref{S_IME} and~\eqref{eMMe_IME} with the most general $2\times 2$ complex hermitian matrix
\begin{equation}
G^{(2)}_{\mathrm{IME}}= \begin{pmatrix}
\Delta_m & \theta_d\mathrm{e}^{\mathrm{-i}\phi_d}\\
\theta_d\mathrm{e}^{\mathrm{i}\phi_d} & \Delta_n
\end{pmatrix}.
\label{G_2_modes}
\end{equation}
Implementing a desired IME thus relies on the capability to engineer $G^{(2)}_\mathrm{IME}$. Its elements represent couplings between two frequency modes. This can be physically implemented by means of systems of coupled cavity, that acts as a frequency beam splitter, and that are accessible to experimental realizations in integrated optics as already demonstrated in the literature~\cite{Hu2021,Lu2018}. 
The approach presented in~\cite{Hu2021}, for example, employs an integrated photonic platform involving two coupled microresonators, as illustrated in \fig{two_mode_meshes}c. 
Through evanescent coupling, the resonances of a pair of hybridized modes can be precisely tuned close to the targeted optical modes $\Omega_m$ and $\Omega_n$, 
achieving the desired elements $\Delta_{m}$ and $\Delta_{n}$.
Additionally, the microwave driving (or electro-optic modulation) applied to the resonators allows to engineer the coupling $\theta\mathrm{e}^{\mathrm{-i}\phi_d}$ between the hybrid modes. 
In particular and $\theta_d$ and $\phi_d$ can be controlled through the driving modulation strength and the phase of the microwave field, respectively.

The complete $N$-mode transformation  $U_{\mathrm{IME}}(\omega)$ can be systematically constructed by choosing a specific order in which 
to perfom the multiplication in~\eqref{smooth decomp} of the appropriate two-mode mode coupled cavity $T_{mn}(\omega)$. 
The frequency-dependent unitary $T_{mn}(\omega)$ is chosen in order to null the element $[U_{\mathrm{IME}}(\omega)]_{mn}$ as described in Appendix~\ref{ap:Smooth_decomp}.
The geometry of the full mesh is defined by the nulling sequence. For example, nulling elements using only other elements of their same column generate a triangular mesh (\fig{two_mode_meshes}a), 
reminiscing of Reck arrangement~\cite{Reck1994}. However such configuration leads to unequal optical path lengths and unbalanced loss due to 
differing numbers of $T_{mn}(\omega)$ operations applied to each mode. 
Alternatively, one can null elements in an order that alternates in a symmetric way between column and row elements, 
to get a much more regular meshing of rectangular shape  (\fig{two_mode_meshes}b), akin to the Clements' scheme \cite{Clements2016}.

Note that, while coupled two-mode cavities serve as useful building blocks, their frequency-dependent coupling profiles may become insufficient to match the required $T_{mn}(\omega)$ over large optical bandwidths.
In particular, as the bandwidth increases, a single two-mode cavity may not have the spectral flexibility needed to implement, even in an approximated way, any arbitrary $T_{mn}(\omega)$ in the whole spectrum. In this condition, the degrees of freedom can be increased by using a linear cascade of two-mode cavities for each  $T_{mn}(\omega)$ or handling each subset of the overall bandwidth with different cavities. 
Such a strategy enables in principle the synthesis of arbitrary $S_{\mathrm{IME}}(\omega)$  over wider, although, of course, at the price of increasing the complexity of the implemented architectures. Again, integrated photonic platforms like lithium niobate (LN) microring resonators are promising choices to implement the full frequency transformation capabilities required by the IME. We show in (\fig{two_mode_meshes}d) and (\fig{two_mode_meshes}e) a possible implementation  in the case of 4 modes, for a triangular and rectangular configuration respectively, in such integrated platforms.

\subsection{Smooth single-mode decomposition}\label{subsec:smooth_1_decomp}
An alternative approach relies on performing the smooth decomposition of $U_{\mathrm{IME}}(\omega)$ in terms of more elementary optical components, involving only 50:50 frequency beam-splitters and single-mode optical cavities.
The advantage is that the only two-mode unitaries are fixed and do not require a specific optimization such as in the approach described in the previous subsection. Although this modular architecture has increased circuit depth, it reduces complex intermodal interactions in our IME to much simpler, well-understood fundamental photonic components building blocks.

By emulating the Reck's and Clements' implementations for spatial modes \cite{Reck1994, Clements2016}, we can rewrite the unitary~\eqref{Tmn} as
\begin{equation}
T_{mn} (\omega)=
\begin{pmatrix} 
\mathrm{e}^{\mathrm{i} \alpha_1(\omega)}& 0 \\ 
0 & \mathrm{e}^{\mathrm{i} \alpha_2(\omega)}
\end{pmatrix}
T_{mn}^{\mathrm{MZI}} (\omega)
\label{u2_omega}
\end{equation}
where the first matrix corresponds to single mode frequency-dependent phase shifters.
On the other side, the unitaries $T_{mn}^{\mathrm{MZI}} (\omega)$ correspond to frequency-dependent Mach-Zehnder interferometers (MZI), as depicted in \fig{meshes}c,
and are given by
\begin{equation}
T_{mn}^{\mathrm{MZI}} (\omega)=
\begin{pmatrix} 
\mathrm{e}^{\mathrm{i} \phi(\omega)}\cos\theta(\omega)& -\sin\theta(\omega) \\ 
\mathrm{e}^{\mathrm{i} \phi(\omega)}\sin\theta(\omega)& \cos\theta(\omega)
\end{pmatrix}.
\end{equation}
The final decomposition therefore writes:
\begin{equation}
U_{\mathrm{IME}}(\omega) = \Phi(\omega) 
\left(
\prod_{(m,n)}T^{\mathrm{MZI}}_{mn}(\omega)\right),
\label{smooth_decomp_2}    
\end{equation}
where in $\Phi(\omega)$ are collected the single mode frequency-dependent phase shifters. These meshes of MZIs can be further decomposed into a series of fixed 50:50 frequency beamsplitters interspersed 
with frequency-dependent single-mode phase shifters (\fig{meshes}c). On the other side, the arbitrary frequency-dependent phase shifter can be implemented by single mode cavities (see \fig{meshes}b), as described in Appendix~\ref{ap:single_mode_cav}. In the case of 4 modes, we show in \fig{meshes}a and \fig{meshes}d two possible implementations of the IME in LN integrated platforms for triangular and rectangular meshes, respectively.

It's opportune to mention again that, as the bandwidth grows, a single cavity becomes insufficient for shaping an arbitrary spectral phase such as 
those that can be encountered in integrated CV quantum photonics~\cite{Gouzien2023}. However, it can always be approximated to any desired accuracy by using a linear chain of single-mode cavities, to either increase the degrees of 
freedom or to let each cavity handle a subset of the overall bandwidth~\cite{Petersen2011, Nurdin2009}.
Therefore, all these arguments allow to formulate the following
\begin{proposition}
Any passive linear transformation $U_{\mathrm{IME}}(\omega)$  between frequency modes can be constructed from fixed 50:50 frequency beamsplitters and chains of 
single-mode cavities approximating frequency phase shifters.
\end{proposition}

\section{Measuring morphing supermodes}\label{sec:mesure MS}
In this section we illustrate an application of IMEs to the field measurement of the non-classical properties of the light generated by single-mode, two-mode and four-mode OPOs. The detection scheme employs an IME before the HD as illustrated in \fig{1_2_mode_detections}(b) and~\fig{1_2_mode_detections}(d).
The analysis of these simple few-mode scenarios elucidates the core capabilities of IMEs and concretely demonstrates its advantages over standard HD.

\subsection{Single-mode OPO}
\begin{figure}[t!]
  \centering
  \includegraphics[width=\linewidth]{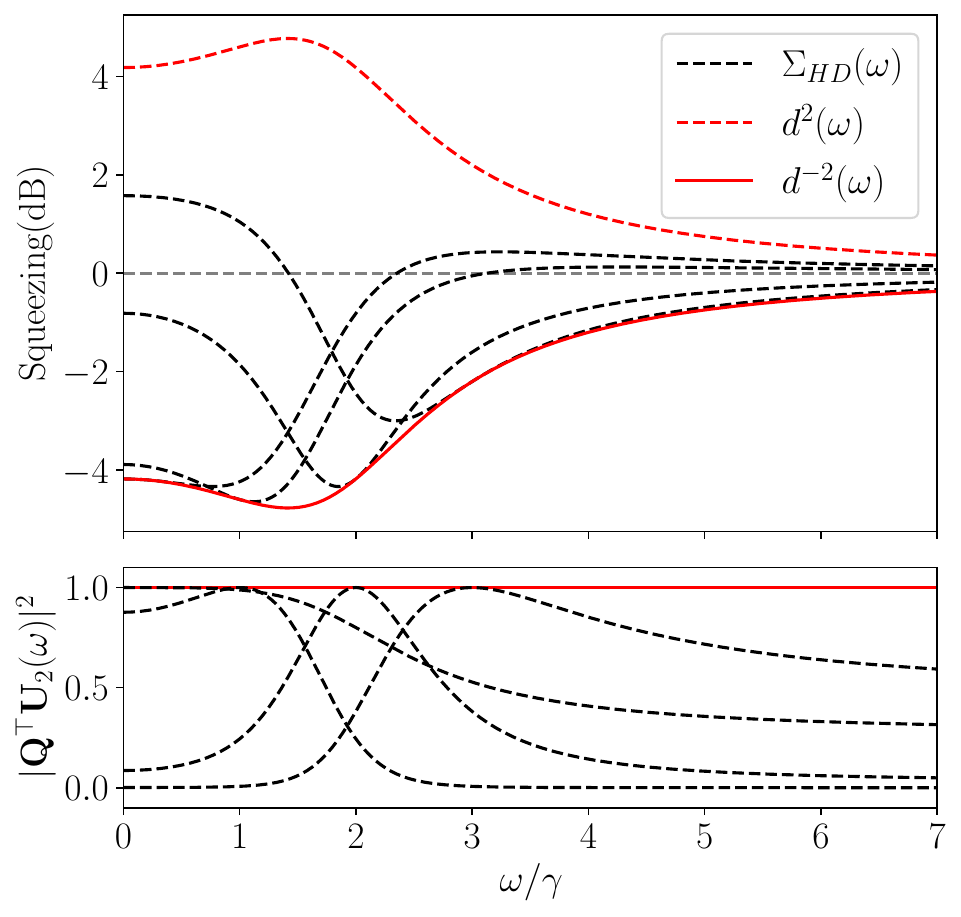}
  \caption{Single-mode OPO: (\textit{top}) comparison between the optimal squeezed and anti-squeezed noise spectral power as predicted by the singular values $d^{-2}(\omega)$ (solid-red line) and $d^{2}(\omega)$ (dashed-red line) and the noise spectral power $\Sigma_{\mathrm{HD}}(\omega)$ as measured through a standard HD for four different choices of $\bm{Q}$ that optimize the detected squeezed signal at a particular $\omega$ (dashed-black lines). The standard quantum limit is the zero dB level. (\textit{Bottom}) Overlapping between a standard LO and the squeezed morphing supermode for four different angles $\theta_{\mathrm{LO}}$ (dashed-black lines) as a function of $\omega$. The solid-red line represents perfect overlapping at all frequencies}
  \label{fig:HD_single_mode_OPO}
\end{figure}

The single-mode degenerate optical parametric oscillator (OPO) represents one of the simplest systems exhibiting squeezing. Here, it will serves as an introductory example to illustrate the limitations of standard HD as well as the advantages of the generalized mode-matching through an IME. 
The system consists of a single-mode nonlinear cavity driven by a continuous wave pump at frequency $\omega_{p}$ and working in frequency and polarization degeneracy, below the OPO oscillation threshold. 
The pump generates, by parametric down-conversion of coupling strength $g$, pairs of signal photons, whose frequency $\omega_{\mathrm{s}}$ is taken to be close to a given resonance $\omega_0$ of the cavity. We will indicate the detuning from perfect resonance as $\Delta = \omega_{\mathrm{s}}-\omega_{0}$ and the signal damping rate as $\gamma$. The dynamics of the signal field can be linearized around stationary values leading to quantum Langevin equations of the form of~\eqref{langevin quad} in the quadrature basis. As we are dealing with a single mode system, the quadrature vector is $\hat{\bm{R}}=(\hat{x}|\hat{y})^{\transp}$, $G=\Delta$ and $F=i g$. The corresponding mode coupling matrix $\mathcal{M}$ is, then, given by
\begin{align} \label{single_mode_opo}
\mathcal{M}= 
\begin{pmatrix}
g & \Delta \\ 
-\Delta & -g
\end{pmatrix}.
\end{align}
The ABMD of the transfer function~\eqref{S} of the system returns two frequency-dependent singular values $d(\omega)$ and $d^{-1}(\omega)$ [see figure \ref{fig:HD_single_mode_OPO}.(\textit{top})], 
representing respectively the optimal levels of anti-squeezing (red-dashed) and squeezing (red-solid). They are associated with two morphing supermodes whose structure is given by the columns of the matrix $U(\omega)$
\begin{align} \label{sm_morphing supermodes}
U(\omega)= 
\begin{pmatrix}
\cos \left[\theta(\omega)\right] & \sin \left[\theta(\omega)\right] \\ 
-\sin \left[\theta(\omega)\right] & \cos \left[\theta(\omega)\right]
\end{pmatrix}.
\end{align}
Since $U(\omega)$ is real, the coefficients of the linear combinations leading to the supermodes are real too. Therefore this example represents a situation where a morphing behaviour is present, but not the hidden squeezing. In what follows, we will focus on the optimal squeezing spectrum $d^{-1}(\omega)$ of the quadrature
$\hat{y}^ {(s)}_{\mathrm{out}}(\omega)=\sin\left[\theta(\omega)\right] \hat{x}_{\mathrm{out}}(\omega)+\cos\left[\theta(\omega)\right] \hat{y}_{\mathrm{out}}(\omega)$. The coefficients of this linear combination are given by $\bm{U}_2(\omega)$, \textit{i.e.} the second column of $U(\omega)$. 
We start by considering the measurement of squeezing through a standard HD, as depicted in \fig{1_2_mode_detections}a, corresponding to a LO with a real and non-morphing profile $\bm{Q}= (\cos \theta_{\mathrm{LO}}|\sin \theta_{\mathrm{LO}})^\transp$. The measured noise spectral power $\Sigma_{\mathrm{HD}}(\omega)$ is traced in \fig{HD_single_mode_OPO}.(\textit{top}) (see black-dashed lines). In the figure, four different choices of $\theta_{\mathrm{LO}}$, thus of $\bm{Q}$, are considered. As it can be seen, for each choice of $\bm{Q}$, optimal squeezing is detected only at a given frequency $\bar{\omega}$ and not through the whole bandwidth: the value of $\bar{\omega}$ corresponds to the frequency at which the LO profile $\bm{Q}$ matches the supermode $\bm{U}_2(\omega)$ and it is obtained by setting 
$\theta_{\mathrm{LO}}=\pi/2-\theta(\bar{\omega})$.
In \fig{HD_single_mode_OPO}.(\textit{bottom}), the sub-optimal mode-matching (black-dashed) between the LO and the morphing supermode $\bm{U}_2(\omega)$ is traced: 
perfect projection is reached only at a fixed frequency. In experiments, this means that the only way to perform a complete reconstruction of squeezing spectrum would be to adjust for any value of   $\omega$, the LO phase so as to reach the minimize the noise spectral power: a procedure highly impractical from the experimental point of view and not at all applicable to time homodyne schemes.
\begin{figure}[t!]
  \centering
  \includegraphics[width=\linewidth]{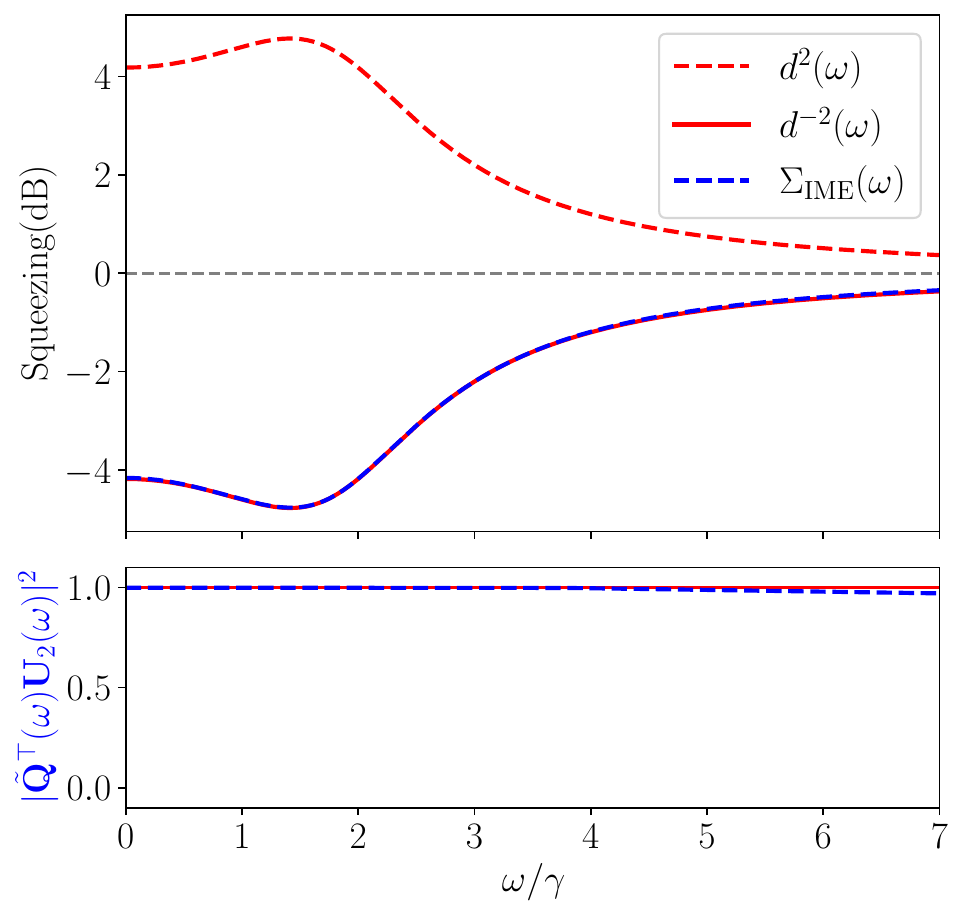}
  \caption{Single-mode OPO: (\textit{top}) comparison between the optimal squeezed and anti-squeezed noise spectral power as predicted by the singular values $d^{-2}(\omega)$ (solid-red line) and $d^{2}(\omega)$ (dashed-red line) and the noise spectral power $\Sigma_{\mathrm{IME}}(\omega)$as measured through IME (dashed-blue line). The standard quantum limit is the zero dB level. (\textit{Bottom}) Overlapping between the generalized LO and the squeezed morphing supermode (dashed-blue line) as a function of $\omega$. The solid-red line represents perfect overlapping at all frequencies.}
  \label{fig:RD_single_mode_OPO}
\end{figure}

This aspect of the HD does not represent, at least in this situation, a limit for the complete reconstruction of the squeezing spectrum that can, indeed, be obtained by sweeping the LO angle. 
The use of IME as interface between the state generation stage and the HD allows to overcome this limitation. 

In order to implement the generalized mode-matching, the cavity output is now made pass through an IME before the HD stage as shown in
\fig{1_2_mode_detections}b.
For single-mode systems, the most general passive transformation $S_{\mathrm{IME}}(\omega)$ (see~\eqref{S_IME}) is characterized by a $2\times 2$ damping matrix
$\Gamma_{\mathrm{IME}}=\mathrm{diag}\{\gamma_{\mathrm{IME}}|\gamma_{\mathrm{IME}}\}$ and $2\times 2$ mode coupling matrix (see~\eqref{eMMe_IME}):
\begin{align} \label{M_interferometer_sm}
\mathcal{M}^{(1)}_{\mathrm{IME}}= 
\begin{pmatrix}
0 & \Delta_{\mathrm{IME}} \\ 
-\Delta_{\mathrm{IME}} & 0
\end{pmatrix}.
\end{align}
The corresponding transfer function is, therefore, the transformation induced by a single mode empty cavity with damping $\gamma_{\mathrm{IME}}$ and detuned with respect to the input beam of a quantity $\Delta_{\mathrm{IME}}$.
As already discussed, such type of "resonator detection'' was proposed by Barbosa \textit{et al.}~\cite{Barbosa2013,BarbosaPRA2013} as a solution to overcome the HD limits in detecting hidden squeezing but never thought as a solution to mode-match a morphing behaviour.

For given values of the parameters of the OPO model ($\Delta=2\gamma$, $g=\gamma$), we can numerically optimize the IME parameters ($\gamma_{\mathrm{IME}}=2\gamma$, $\Delta_{\mathrm{IME}}=-1.51\gamma$) and the LO angle ($\theta_{\mathrm{LO}}=4.96$ rad) in order to achieve optimal mode-matching over the largest bandwidth. 
These results are shown in \fig{RD_single_mode_OPO}.(\textit{top}), where the dashed-blue curve represents the noise spectral power detected by a standard HD after 
a passage of the signal through the IME. In \fig{RD_single_mode_OPO}.(\textit{bottom}) we show that the projection between the generalized LO, $\tilde{\bm{Q}}(\omega)$, and the squeezed supermode is optimal (close to one) through the whole bandwith. As we will see next, this advantage becomes even stronger in multimode scenarios.

\subsection{Two-mode OPO}
\begin{figure}[h!]
\centering
\includegraphics[width=1\linewidth]{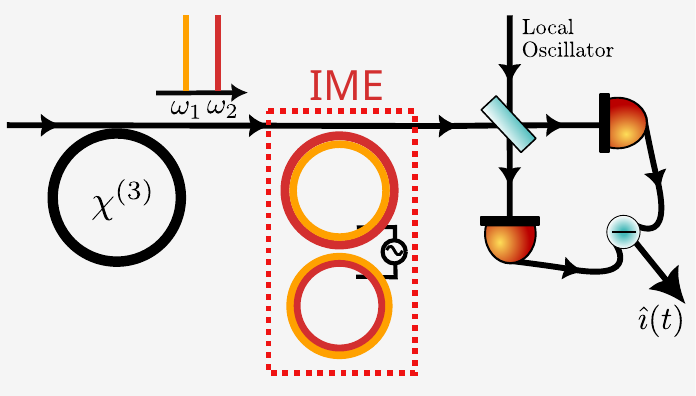}
\caption{Schematic of a generalized mode matching between a bi-color quantum field from a $\chi^{(3)}$ OPO source and the standard HD,  using a coupled cavity system as an IME.}
\label{fig:IME_2modes}
\end{figure}
\begin{figure}[t!]
  \centering
  \includegraphics[width=\linewidth]{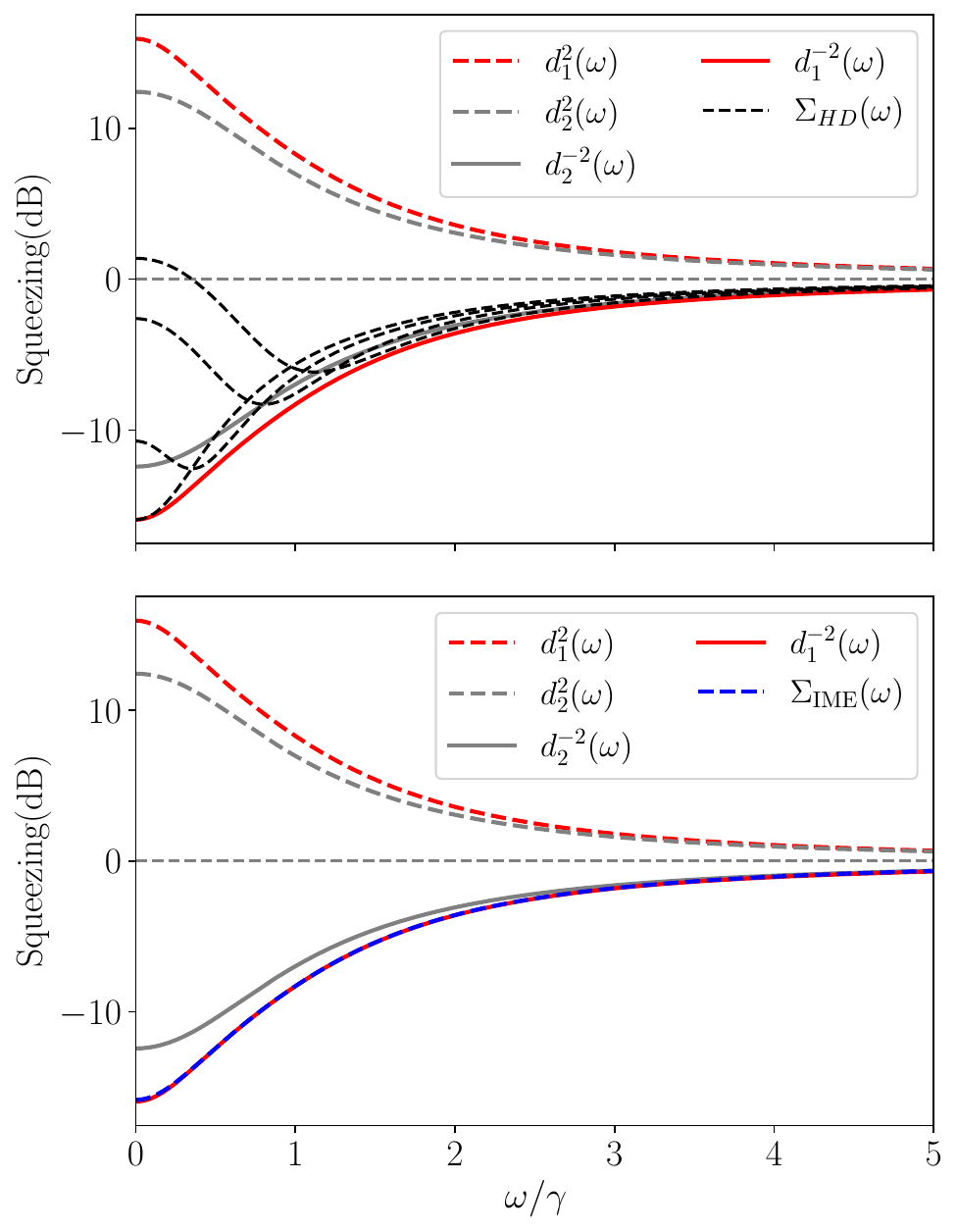}
  \caption{Two-mode OPO: (\textit{top}) comparison between noise spectral power of the most squeezed and anti-squeezed morphing supermodes as predicted by the singular values $d_1^{-2}(\omega)$ (solid-red line) and $d_1^{2}(\omega)$ (dashed-red line) and the noise spectral power $\Sigma_{\mathrm{HD}}(\omega)$ as measured through a standard HD for four different choices of $\bm{Q}$ that optimize the detected squeezed signal at a particular $\omega$ (dashed-black lines). The solid and dashed gray lines represent the noise of the least squeezed and anti-squeezed morphing supermodes. The standard quantum limit is the zero dB level. (\textit{Bottom}) The noise spectral power $d_1^{-2}(\omega)$ of the most squeezed morphing supermodes is fully recovered by inserting an IME before the HD.}
  \label{fig:two_modes_OPO}
\end{figure}

The case of two-mode OPO clearly shows the limitations of HD to fully characterize the quantum dynamics, even in simple bimodal scenarios.
We consider a non degenerate OPO, pumped in continuous wave below the oscillation threshold. This case is described by a system of linear quantum Langevin equations~\eqref{langevin quad} with 
$4\times 4$ damping matrix $\Gamma=\mathrm{diag}\{\gamma_{1},\gamma_{2}|\gamma_{1},\gamma_{2}\}$ and $4\times 4$ mode-coupling matrices~\eqref{eMMe} corresponding to
\begin{align} \label{two_modes_FG}
G = \begin{pmatrix}
g_{11} & g_{12} \\ 
g_{12}^{\ast} & g_{22}
\end{pmatrix},\;
F = \begin{pmatrix}
0 & f_{12} \\ 
f_{12} & 0
\end{pmatrix}.
\end{align}
Despite the apparent simplicity of this two-mode system, the interplay between parametric amplification and dispersion leads to hidden squeezing and complex morphing supermode structure~\cite{Dioum2024}.
The noise spectral power measured by a traditional HD detection scheme with a two-color LO, as depicted in \fig{1_2_mode_detections}c, is not even able reach the optimal squeezing value of the first supermode [respectively, the black-dashed and red lines in \fig{two_modes_OPO}-(\textit{top})]
as like in the example with a single-mode OPO.
This is due to a sub-optimal mode-matching problem between the LO, associated to a real $\bm{Q}$, and a complex supermode structure $U(\omega)$.
The advantage of using IME with two-mode fields is even more straiking then the single-mode case since it allows the characterization 
of quantum correlations that are inaccessible with conventional HD. Instead, a generalized mode-matching that produces a complex frequency-dependent LO is necessary (see Section~\ref{sec:generalized mode-matching}).
The most general passive transformation $S_{\mathrm{IME}}(\omega)$ [see~\eqref{S_IME}] is characterized by a $4\times 4$ damping matrix
$\Gamma_{\mathrm{IME}}=\mathrm{diag}\{\gamma_{\mathrm{\tiny IME},1},\gamma_{\mathrm{\tiny IME},2}|\gamma_{\mathrm{\tiny IME},1},\gamma_{\mathrm{\tiny IME},2}\}$ and $4\times 4$ mode coupling matrix~\eqref{eMMe_IME} 
with a $G_{\mathrm{IME}}$ given by the most general Hermitian matrix~\eqref{G_2_modes}. Therefore, as explained in Sectionref{sec:smooth dec}.\ref{subsec:Smooth_2_decomp}, 
this two-mode IME can be implemented by the coupled cavity system depicted in figure \ref{fig:IME_2modes}, that is the main building block of our smooth two-mode decompositions. 

For given values of the two-mode OPO ($\gamma_1$, $\gamma_2=\gamma_1$, $g_{11}=0.8\gamma_1$, $g_{12}=0.1\gamma_1$, $g_{22}=\gamma_1$, $f_{12}=i\gamma_1$), the goal is to optimize the IME parameters so to shape the generalized LO, $\tilde{\bm{Q}}(\omega)$, as the the most squeezed morphing supermode, the structure of which is given by the $(N+1)$-th column of $U(\omega)$. We found that the use of just one IME allows to implement a generalized LO allowing mode-matching in the first half of the spectrum. In order to achieve optimal matching over the full spectrum, we employed two chained IME systems. In \fig{two_modes_OPO}.(\textit{bottom}) 
we show that the noise spectrum $d_1^{-2}(\omega)$ of the most squeezed morphing supermode can be completely recovered by setting the parameters as 
$\gamma_{\mathrm{\tiny IME},1}=\gamma_{\mathrm{\tiny IME},2}=0.91\gamma_1$, $\Delta_1=-4.59\gamma_1$, $\Delta_2=-2.86\gamma_1$, $\theta_d=3.67\gamma_1$ and $\phi_d=10.77$ rad for the first IME and 
$\gamma_{\mathrm{\tiny IME},1}=\gamma_{\mathrm{\tiny IME},2}=1.45\gamma_1$, $\Delta_1=-2.42\gamma_1$, $\Delta_2=-1.62\gamma_1$, $\theta_d=1.43\gamma_1$ and $\phi_d=12.6$ rad for the second IME. 
The optimization also involves the three angles of the the two-color LO, characterized by 
\begin{align*}
\bm{Q}=\big(\bm{x}_{\mathrm{LO}}
\big|
\bm{y}_{\mathrm{LO}}
\big)^\transp,
\end{align*}
where 
\begin{align*}
\bm{x}_{\mathrm{LO}}
&=
\Big(
\cos\theta_{\mathrm{LO},1}\cos\theta_{\mathrm{LO},2}
\;,\;
\cos\theta_{\mathrm{LO},3}\sin\theta_{\mathrm{LO},2}
\Big)^\transp
\end{align*}
and
\begin{align*}
\bm{y}_{\mathrm{LO}}
&=
\Big(
\sin\theta_{\mathrm{LO},1}\cos\theta_{\mathrm{LO},2}
\;,\;
\sin\theta_{\mathrm{LO},3}\sin\theta_{\mathrm{LO},2}
\Big)^\transp,
\end{align*}
and it returns $\theta_{\mathrm{LO},1}=10.44$ rad, $\theta_{\mathrm{LO},2}=1.47$ rad and $\theta_{\mathrm{LO},3}=7.69$ rad.
This example showcases the substantial enhancement provided by the IME approach
that allows to mode-match complex morphing supermodes thus allowing to access to correlations that are hidden to standard HD.

\subsection{Four-mode OPO}
\begin{figure}[t!]
  \centering
  \includegraphics[width=\linewidth]{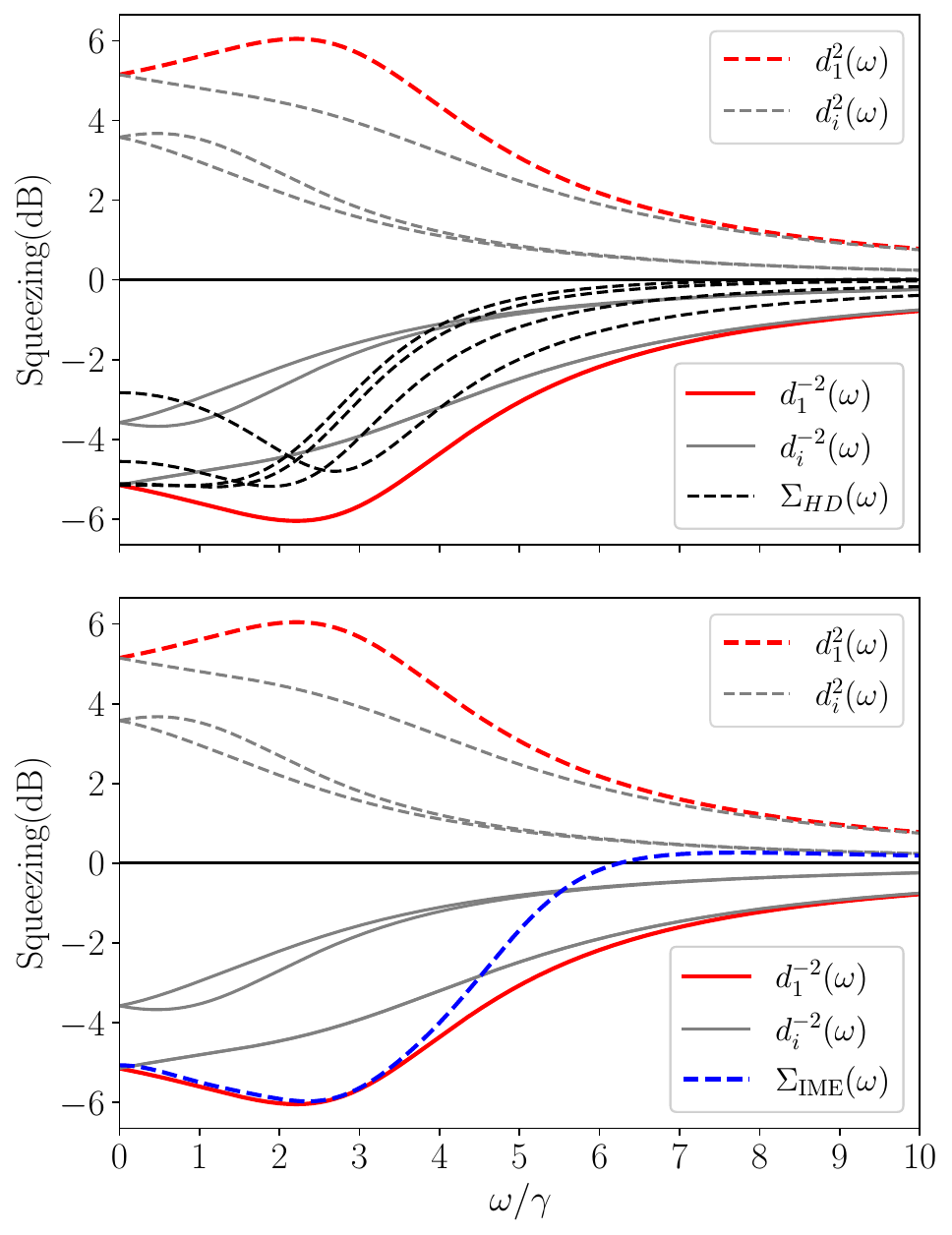}
  \caption{Four-mode OPO: (\textit{top}) comparison between noise spectral power of the most squeezed and anti-squeezed morphing supermodes as predicted by the singular values $d_1^{-2}(\omega)$ (solid-red line) and $d_1^{2}(\omega)$ (dashed-red line) and the noise spectral power $\Sigma_{\mathrm{HD}}(\omega)$ as measured through a standard HD for four different $\theta_{\mathrm{LO}}$ that optimize the detected squeezed signal at a particular $\omega$ (dashed-black lines). The solid and dashed gray lines represent the noise $d_i^{-2}(\omega)$ of the least squeezed and $d_i^{-2}(\omega)$ of the least anti-squeezed morphing supermodes (for $i=\{2,3,4\}$). The standard quantum limit is the zero dB level. (\textit{Bottom}) The noise spectral power $d_1^{-2}(\omega)$ of the most squeezed morphing supermodes is fully recovered by inserting an IME before the HD.}
  \label{fig:four_modes_OPO}
\end{figure}

In order to demonstrate the scalability of the IME, we consider a four-mode quadratic open quantum system with dampings 
$\{\gamma_1,\gamma_2,\gamma_3,\gamma_4\}$ described by the following interaction matrices:
\begin{equation} \label{four_modes}
G=
\left(
\begin{array}{cccc}
2a & 0 & a & 0
\\
0 & 2a & 0 & a
\\
a & 0 & 2a & 0
\\
0 & a & 0 & 2a
\end{array}
\right),\;
F=
\left(
\begin{array}{cccc}
0 & b & 0 & 2b
\\
b & 0 & 2b & 0
\\
0 & 2b & 0 & b
\\
2b & 0 & b & 0
\end{array}
\right)
\end{equation}
and set the parameters as following: $\gamma_2=1.5\gamma_1$, $\gamma_3=\gamma_1$, $\gamma_4=1.5\gamma_1$, $a=0.55\gamma_1$ and $b=0.3\gamma_1$. 
This four-mode system could describe, for example, a $\chi^{(3)}$ nonlinear cavity driven by strong pumps that lead to parametric amplification and 
cross-phase modulation couplings between the modes~\cite{Bensemhoun2024}. The ABMD allows to obtain the morphing supermodes (that are complex in this example) 
and their corresponding level of squeezing or anti-squeezing. In particular, four supermodes present squeezed spectra $d_i^{-1}(\omega)$ (for $i=1,\ldots,4$) and 
the others present anti-squeezed spectra $d_i(\omega)$). In \fig{four_modes_OPO}, the red-solid curve represents the squeezing spectrum of the most squeezed morphing supermode, 
the red-dashed is the noise spectrum of the most anti-squeezed morphing supermode, while the grey-solid and grey-dashed curves represent the noise spectra of the other supermodes.
In \fig{four_modes_OPO}.(\textit{top}), the black-dashed curves represent the noise spectral power detectable by a standard HD (with a four-color LO). 
Each spectrum is obtained for a specific LO optimized in order to measure the maximal squeezing possible at a given frequency. Because of the impossibility of 
reaching perfect mode-matching, HD fails to recover the noise spectrum $d_1^{-1}(\omega)$ of the most squeezed morphing supermode (red-solid curve) at all frequencies 
but $\omega=0$. Here, the spectral covariance matrix and the morphing supermodes are always real as it can be deduced from~\eqref{S}. 

In contrast, optimal mode-matching can be achieved by introducing a four-mode IME between the source and the HD as depicted, for example, in the case of a rectangular mesh in \fig{1_2_mode_detections}d. Its transfer function $S_{\mathrm{IME}}(\omega)$ [\eqref{S_IME}] 
is now described by a $8\times 8$ matrix with 17 free parameters if we choose equal damping rates for all the modes. We use the smooth decomposition developed 
in Sectionref{sec:smooth dec} in order to find a physical implementation for this sytem. As commented above, one can choose either a smooth-triangular or a smooth-rectangular decomposition.
The result of this algorithm gives a structure represented in figs.~\ref{fig:two_mode_meshes}.(a) and~\ref{fig:two_mode_meshes}.(b). Each frequency mode, labelled with the numbers $i=1,\ldots,4$ 
and different colours (yellow, red, green and blue) interacts with the other modes through two-mode frequency beam-splitters (\fig{two_mode_meshes}c). 
These pairwise interactions, represented by black solid squares, are implemented one by one, in the order specified by the figure. 
The physical implementation corresponding to these meshes is, instead, depicted in \fig{two_mode_meshes}d for the case of a smooth-triangular decomposition 
and in \fig{two_mode_meshes}d for the case of a smooth-rectangular decomposition.
In these figures, the two colors of the frequency beam-splitters has been chosen in order to specify which frequency modes they couple.

We can also adopt smooth single-mode decompositions (triangular or rectangular), as described in Sectionref{sec:smooth dec}.\ref{subsec:smooth_1_decomp} in terms of single mode cavities 
and 50:50 frequency beam splitters. The result of this decomposition for the example considered is represented in \fig{meshes}a for 
the triangular and in \fig{meshes}d for the rectangular mesh.

The optimization of the 17 free parameters of the IME transformation allows to mode-match the generalized LO to the targeted supermode: 
here we consider the most squeezed one. We show in \fig{four_modes_OPO}.(\textit{bottom}) that the detected noise spectral power (blue-dashed curve)
overlaps with the desired squeezing spectrum $d_1^{-1}(\omega)$ in the most significant part of the total bandwidth. 
We can chain as many IMEs as necessary to increase the bandwidth where mode-matching is optimal.

\section{Conclusion}
In the framework of CV quantum optics, the most simple and widespread approach for the generation of non-classical states relies on quadratic Hamiltonians and cavity-based dissipative dynamics. Their interplay produces states that present a morphing behaviour and hidden squeezing. These effects, much more common than previously thought, are expected to be present in microring resonators, optomechanics, four-wave mixing (FWM) in atomic ensembles, polaritons in semiconductor microcavities, and quantum cascade lasers. Homodyne detection is insufficient to deal with these states, this hinders the efficiency of CV protocols for QIP. We showed in this paper that this is due to the impossibility of HD of realizing perfect mode-match even in the best scenario. We presented a universal approach in order to obtain perfect mode-matching through a novel detection scheme based on interferometers with memory effect. We finally designed a physical implementation of these devices. These results promise to unlock the potentialities of integrated quantum photonics for applications in QIP.

\begin{backmatter}
\bmsection{Funding} We acknowledge funding from the Plan France 2030 through the project ANR-22-PETQ-0013. Virginia D'Auria acknowledges financial support from the Institut Universitaire de France (IUF) and the project SPHIFA (ANR-20-CE47-0012). Olivier Pfister acknowledges financial support from U.S. National Science Foundation grants PHY-2112867 and ECCS-2219760. Giuseppe Patera acknowledges financial support from the Agence Nationale de la Recherche through LABEX CEMPI (ANR-11-LABX-0007) and I-SITE ULNE (ANR-16-IDEX-0004).

\bmsection{Acknowledgments} We are grateful for discussions to \'{E}lie Gouzien about interferometers with memory effect, to Carlos Navarrete-Benlloch about the spectral covariance matrix and to Alessandro Pugliese about analytic decompositions.

\bmsection{Disclosures} The authors declare no conflicts of interest.

\bmsection{Data availability} No data were generated or analyzed in the presented research.

\bmsection{Supplemental document}
See Supplement 1 for supporting content. 

\end{backmatter}

\section*{Appendices}
\appendix
\section{Positivity of the noise spectral power}\label{ap:NSP}
The ABMD of a $2N\times2N$ Hermitian spectral covariance matrix gives
\begin{align}
\sigma(\omega)&=U(\omega) D(\omega)^2 U^\dagger(\omega) 
\end{align}
where $U(\omega)$ and $D(\omega)$ are the smooth matrix-valued functions of $\omega$ detailed in the main text after~\eqref{ABMD}. For sake of simplicity we will omit the dependence on $\omega$ in the following expressions. Since $U$ is complex, it can be written in terms of its real and imaginary parts $U_{\mathrm{re}}+i U_{\mathrm{im}}$. Then, the noise spectral power~\eqref{noisespectrum} is
\begin{align}
\Sigma_{\bm{Q}}
&=
\bm{Q}^\transp
\left(U_{\mathrm{re}} D^2 U_{\mathrm{re}}^\transp + U_{\mathrm{im}} D^2 U_{\mathrm{im}}^\transp\right)
\bm{Q}
+
\nonumber
\\
&+i\bm{Q}^\transp
\left(U_{\mathrm{im}} D^2 U_{\mathrm{re}}^\transp - U_{\mathrm{re}} D^2 U_{\mathrm{im}}^\transp\right)
\bm{Q}.
\end{align}
Thus, the reality of $\Sigma_{\bm{Q}}$ is proven by showing that 
$\bm{Q}_{\mathrm{im}}^\transp D^2 \bm{Q}_{\mathrm{re}}=\bm{Q}_{\mathrm{re}}^\transp D^2 \bm{Q}_{\mathrm{im}}$, where we defined the column vectors
$\bm{Q}_{\mathrm{re}}=U_{\mathrm{re}}^\transp\bm{Q}$ and $\bm{Q}_{\mathrm{im}}=U_{\mathrm{im}}^\transp\bm{Q}$. This is indeed the case, since:
\begin{align}
\bm{Q}_{\mathrm{im}}^\transp D^2 \bm{Q}_{\mathrm{re}}
&=\sum_{k} \left(\bm{Q}_{\mathrm{im}}\right)_k \left(\bm{Q}_{\mathrm{re}}\right)_k d_k^2
=\bm{Q}_{\mathrm{re}}^\transp D^2 \bm{Q}_{\mathrm{im}}.
\end{align}

\section{Smooth Decompositions}\label{ap:Smooth_decomp}

As we introduced in Sectionref{sec:smooth dec}.\ref{subsec:Smooth_2_decomp}, any unitary $U{_\mathrm{IME}}(\omega)$ can be decomposed into a product of at most $N(N-1)/2$ 
two-mode frequency-dependent unitaries $T_{mn}(\omega)$ through~\eqref{smooth decomp}. Here we describe more in details this novel type of decomposition and the felxibility in the decomposition method.

Particularly, the $T_{m,n}(\omega)$, implemented through coupled cavities, are $N\times N$ unitary transformations involving the modes $m$ and $n$, 
leaving the $N-2$ modes unchanged. If we choose for example a practical situation where the decomposition involves only neighbouring modes $m$ and $n = m \pm 1$, 
the complete form if $T_{mn}(\omega)$ have the form :
\begin{equation}
T_{mn} (\omega)=
\begin{pmatrix}
1 & 0 & \cdots & \cdots & \cdots & \cdots & \cdots& 0\\ 
0 & 1 &  &  &  &  &   & \vdots \\ 
\vdots &  & \ddots  &  &   &  &  & \vdots\\ 
\vdots &  &  & \mathrm{e}^{\mathrm{i} \phi(\omega)}a(\omega)& -b(\omega) &  & \\ 
\vdots &  &  & \mathrm{e}^{\mathrm{i} \phi(\omega)}b^{\ast}(\omega)& a^{\ast}(\omega)&  & \\ 
\vdots &  &  &  &  & \ddots &  &\vdots \\ 
\vdots &  &  &  &  &  & 1 & 0\\ 
0 & \cdots & \cdots & \cdots & \cdots & \cdots  & 0 &1 
\end{pmatrix},
\label{eq:Tmn_omega}    
\end{equation}
where $a(\omega)$ and $b(\omega)$ are frequency dependant complex coefficients satisfying $|a(\omega)|^{2}+|b(\omega)|^{2}=1$ and $\mathrm{e}^{\mathrm{i} \phi(\omega)}$ the determinant of the two level unitary.  
In the main text [\eqref{Tmn}], we have assimilited the $T_{mn} (\omega)$ to their form in their most interesting two-dimension subspace.

The flexibility of this novel decomposition lays in the choice of these building block $T_{m,n}(\omega)$.
Such choice depends on the decomposition order and the allowed structure of the unitaries provided the product of determinants satisfies:
\begin{align} \label{det_Tmn}
\prod_{(mn)}\mathrm{det} \Big[T_{m,n}
(\omega)\Big] =
\mathrm{det} [U(\omega)].
\end{align}
This decomposition relies on extending the concepts in \cite{Li2013} to unitary matrices smooth in continuous variable parameters, like frequency. 
Specifically, such decomposition is possible because:
\begin{lemma}
By using matrix elements in the same row or column, any element $u_{mn}(\omega)$ of a unitary matrix $U_{\mathrm{IME}}(\omega)$ can be nulled through 
the multiplication by the appropriate two-level unitaries $T_{m,n}(\omega)$.
\end{lemma}

We illustrate this lemma in the case of a $4 \times 4$ smooth unitary matrix
\begin{align}
U_{\mathrm{IME}}(\omega)=
\begin{bmatrix}
u_{11}(\omega) & u_{12}(\omega) & u_{13}(\omega) & u_{14}(\omega) \\
u_{21}(\omega) & u_{22}(\omega) & u_{23}(\omega) & u_{24}(\omega) \\
u_{31}(\omega) & u_{32}(\omega) & u_{33}(\omega) & u_{34}(\omega) \\
u_{41}(\omega) & u_{42}(\omega) & u_{43}(\omega) & u_{44}(\omega)
\end{bmatrix},
    \label{eq:example_4}
\end{align}
however the general case to $N$ dimensions can be proved by recursion.

To null for example the element $u_{21}(\omega)$ using an element of the same column (say $u_{31}(\omega)$), we can use the following frequency-dependnat transformation between modes 2 and 3:
%
%\begin{widetext}
\begin{align}
&T_{23}(\omega)=\nonumber\\
&{[|u_{21}(\omega)|^{2}+ |u_{31}(\omega)|^2]^{-\tfrac12}}
\begin{bmatrix}
1 & 0 & 0 & 0 \\
0 &  \mathrm{e}^{\mathrm{i}\phi_{23}(\omega)} u_{31}(\omega) & - u_{21}(\omega) & 0 \\
0 & \mathrm{e}^{\mathrm{i}\phi_{23}(\omega)}u_{21}^{\ast}(\omega) & u_{31}^{\ast}(\omega) & 0 \\
0 & 0 & 0 & 1
\end{bmatrix}\\
&\implies
T_{23}(\omega)U(\omega)=
\begin{bmatrix}
* & * & * & * \\
0 &  * & * & * \\
* &  *& * & * \\
* & * & * & * \\
\end{bmatrix}.
\label{eq:delete_column}
\end{align}
%\end{widetext}
%
Similarly, one can choose to null the element $u_{21}(\omega)$ using instead an element of the same line (say $u_{22}(\omega)$) with the following frequency-dependant transformation between modes 1 and 2:
%
%\begin{widetext}
\begin{align}
&T_{12}(\omega)=\\
&\frac{1}{\sqrt{|u_{21}(\omega)|^{2}+ |u_{22}(\omega)|^2}}
\begin{bmatrix}
1 & 0 & 0 & 0 \\
0 &  \mathrm{e}^{\mathrm{i}\phi_{12}(\omega)} u_{22}(\omega) & - u_{21}(\omega) & 0 \\
0 & \mathrm{e}^{\mathrm{i}\phi_{12}(\omega)}u_{21}^{\ast}(\omega) & u_{22}^{\ast}(\omega) & 0 \\
0 & 0 & 0 & 1
\end{bmatrix}\\
& \implies
U(\omega)T^{\dagger}_{12}(\omega)=
\begin{bmatrix}
* & * & * & * \\
0 &  * & * & * \\
* &  *& * & * \\
* & * & * & * \\
\end{bmatrix}.
\label{eq:delete_line}
\end{align}
%\end{widetext}
%

Each nulling operation corresponds to a two-mode coupled cavity between adjacent modes. 
The complete $N$-mode transformation $U_{\mathrm{IME}}(\omega)$ can be systematically constructed by choosing a specific order in which to null its elements, 
which defines the specific order in which to apply the  $T_{m,n}(\omega)$.
\begin{figure*}[ht!]
\centering
\includegraphics[width=\linewidth]{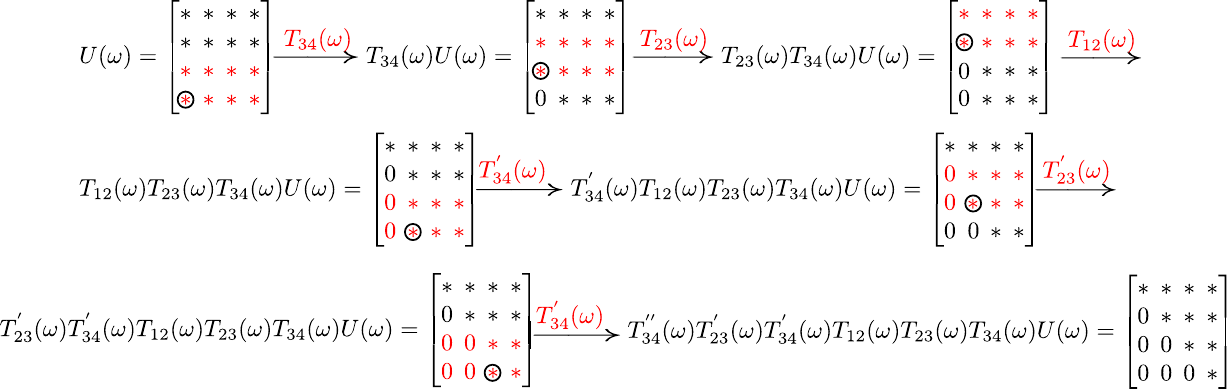}
\caption{Smooth decomposition method for triangular decomposition}
\label{fig:Smooth_decomp_triangular}
\end{figure*}

A conceptually simple approach is to null them column-by-column from the bottom left, going up, as shown in Fig .\ref{fig:Smooth_decomp_triangular} for 4 modes. 
Once all the lower diagonal elements have been nulled, the final product of smooth matrix is a upper triangular matrix. Due to the unitary nature of $U_{\mathrm{IME}}(\omega)$, 
such upper triangular matrix is necessarily diagonal. And by respecting the determinant structure from~\eqref{det_Tmn}, such diagonal matrix is necessarily the identity matrix. 
The final decomposition, for example in the case of four modes :
\begin{equation}
U_{\mathrm{IME}}(\omega)=
T_{34}^{\dagger}(\omega)T_{23}^{\dagger}(\omega)
T_{12}^{\dagger}(\omega)T_{34}^{'\dagger}(\omega)
T_{23}^{'\dagger}(\omega)T_{34}^{''\dagger}(\omega),
\end{equation}
maps onto a triangular mesh connectivity between the frequency modes (\fig{meshes}a), reminiscent of the Reck architecture \cite{Reck1994} for spatial modes.
\begin{figure*}[ht!]
\centering
\includegraphics[width=\linewidth]{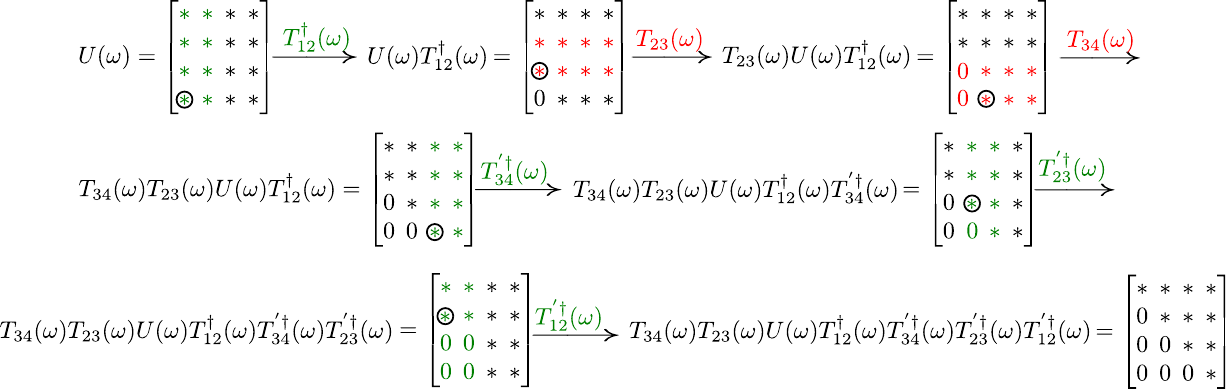}
\caption{Smooth decomposition method for rectangular decomposition}
\label{fig:Smooth_decomp_rectangular}
\end{figure*}

Alternatively, one can null elements in an order that alternates between column and row elements, akin to the Clements scheme~\cite{Clements2016}. 
For example, for 4 modes, the process is shown in \fig{Smooth_decomp_rectangular} and the final decomposition writes:
\begin{equation}
U_{\mathrm{IME}}(\omega)=
T_{23}^{\dagger}(\omega) 
T_{34}^{\dagger}(\omega) T^{'}_{12}(\omega)
T^{'}_{23}(\omega)
T^{'}_{34}(\omega)T_{12}(\omega).
\end{equation}
This gives a highly regular tiling of two-mode frequency coupling, with average equal path lengths between modes (\fig{meshes}b).

\section{Single mode cavity}\label{ap:single_mode_cav}
The quantum Langevin equation of a single-mode single-sided cavity in the interaction picture is:
\begin{equation}
\frac{d}{d t}\hat{a}(t) = (-\gamma + \mathrm{i} \Delta) \hat{a}(t) +\sqrt{2\gamma}\; \hat{a}_{\mathrm{in}}(t),
\label{QL_single annx}
\end{equation}
with $\hat{a}$ the boson annihilation operator, $\gamma$ the cavity damping, $\Delta$ the detuning from cavity resonance and  $\hat{a}_{in}$ the input mode entering the cavity via losses. 
In the Fourier space, we get:
\begin{equation}
\hat{a}(\omega) = \frac{\sqrt{2\gamma}}{\gamma -\mathrm{i} (\omega+\Delta)}\; \hat{a}_{\mathrm{in}}(\omega),
\label{Fourier annx }
\end{equation}
and using the input-output relations $\hat{a}_{\mathrm{out}}(t)+\hat{a}_{\mathrm{in}}(t)=\sqrt{2\gamma}$, we get the output mode:
\begin{equation}
\hat{a}_{\mathrm{out}}(\omega) = \frac{\gamma+ \mathrm{i}(\omega + \Delta)}{\gamma -\mathrm{i} (\omega+\Delta)}\; \hat{a}_{\mathrm{in}}(\omega),
\label{single_in_out annex}
\end{equation}
which is an $\omega$-dependant unitary transformation equivalent to a $\omega$ dependant phase shift:
\begin{equation}
\hat{a}_{\mathrm{out}}(\omega) = \mathrm{e}^{i\theta(\omega)} \hat{a}_{\mathrm{in}}(\omega).
\label{frequency_phase_shift}
\end{equation}

\FloatBarrier
%%%%%%%%%%%%%%%%%%%%%%%%%%%%%%%%%%%%%%%%%%%%%%%%%%
%\bibliography{biblio,Pfister}

\begin{thebibliography}{10}
\newcommand{\enquote}[1]{``#1''}

\bibitem{Pfister2024}
O.~Pfister, \enquote{Qubits without qubits,} {\protect\JournalTitle{Science}}
  \textbf{383}, 264--264 (2024).

\bibitem{Gottesman2001}
D.~Gottesman, A.~Kitaev, and J.~Preskill, \enquote{Encoding a qubit in an
  oscillator,} {\protect\JournalTitle{Phys. Rev. A}} \textbf{64}, 012310
  (2001).

\bibitem{Menicucci2014ft}
N.~C. Menicucci, \enquote{Fault-tolerant measurement-based quantum computing
  with continuous-variable cluster states,} {\protect\JournalTitle{Phys. Rev.
  Lett.}} \textbf{112}, 120504 (2014).

\bibitem{Chen2014}
M.~Chen, N.~C. Menicucci, and O.~Pfister, \enquote{Experimental realization of
  multipartite entanglement of 60 modes of a quantum optical frequency comb,}
  {\protect\JournalTitle{Phys. Rev. Lett.}} \textbf{112}, 120505 (2014).

\bibitem{Yokoyama2013}
S.~Yokoyama, R.~Ukai, S.~C. Armstrong, \emph{et~al.},
  \enquote{Ultra-large-scale continuous-variable cluster states multiplexed in
  the time domain,} {\protect\JournalTitle{Nat. Photon.}} \textbf{7}, 982
  (2013).

\bibitem{Yoshikawa2016}
J.-i. Yoshikawa, S.~Yokoyama, T.~Kaji, \emph{et~al.}, \enquote{Invited article:
  Generation of one-million-mode continuous-variable cluster state by unlimited
  time-domain multiplexing,} {\protect\JournalTitle{APL Photonics}} \textbf{1},
  060801 (2016).

\bibitem{Asavanant2019}
W.~Asavanant, Y.~Shiozawa, S.~Yokoyama, \emph{et~al.}, \enquote{Generation of
  time-domain-multiplexed two-dimensional cluster state,}
  {\protect\JournalTitle{Science}} \textbf{366}, 373--376 (2019).

\bibitem{Larsen2019}
M.~V. Larsen, X.~Guo, C.~R. Breum, \emph{et~al.}, \enquote{Deterministic
  generation of a two-dimensional cluster state,}
  {\protect\JournalTitle{Science}} \textbf{366}, 369--372 (2019).

\bibitem{Lita2008}
A.~E. Lita, A.~J. Miller, and S.~W. Nam, \enquote{Counting near-infrared
  single-photons with 95\% efficiency,} {\protect\JournalTitle{Opt. Expr.}}
  \textbf{16}, 3032--3040 (2008).

\bibitem{Eaton2023}
M.~Eaton, A.~Hossameldin, R.~J. Birrittella, \emph{et~al.}, \enquote{Resolution
  of 100 photons and quantum generation of unbiased random numbers,}
  {\protect\JournalTitle{Nature Photonics}} \textbf{17}, 106--111 (2023).

\bibitem{Cahall2017}
C.~Cahall, K.~L. Nicolich, N.~T. Islam, \emph{et~al.}, \enquote{Multi-photon
  detection using a conventional superconducting nanowire single-photon
  detector,} {\protect\JournalTitle{Optica}} \textbf{4}, 1534--1535 (2017).

\bibitem{Wallentowitz1996}
S.~Wallentowitz and W.~Vogel, \enquote{Unbalanced homodyning for quantum state
  measurements,} {\protect\JournalTitle{Phys. Rev. A}} \textbf{53}, 4528--4533
  (1996).

\bibitem{Banaszek1996}
K.~Banaszek and K.~{W\'odkiewicz}, \enquote{Direct probing of quantum phase
  space by photon counting,} {\protect\JournalTitle{Phys. Rev. Lett.}}
  \textbf{76}, 4344 (1996).

\bibitem{Sridhar2014a}
N.~Sridhar, R.~Shahrokhshahi, A.~J. Miller, \emph{et~al.}, \enquote{Direct
  measurement of the {W}igner function by photon-number-resolving detection,}
  {\protect\JournalTitle{J. Opt. Soc. Am. B}} \textbf{31}, B34--B40 (2014).

\bibitem{Nehra2019}
R.~Nehra, A.~Win, M.~Eaton, \emph{et~al.}, \enquote{State-independent quantum
  state tomography by photon-number-resolving measurements,}
  {\protect\JournalTitle{Optica}} \textbf{6}, 1356--1360 (2019).

\bibitem{Olivares2019}
S.~Olivares, A.~Allevi, G.~Caiazzo, \emph{et~al.}, \enquote{Quantum tomography
  of light states by photon-number-resolving detectors,}
  {\protect\JournalTitle{New Journal of Physics}} \textbf{21}, 103045 (2019).

\bibitem{Nehra2020a}
R.~Nehra, M.~Eaton, C.~Gonz\'alez-Arciniegas, \emph{et~al.},
  \enquote{Generalized overlap quantum state tomography,}
  {\protect\JournalTitle{Phys. Rev. Research}} \textbf{2}, 042002 (2020).

\bibitem{Lvovsky2013}
A.~I. Lvovsky, R.~Ghobadi, A.~Chandra, \emph{et~al.}, \enquote{Observation of
  micro--macro entanglement of light,} {\protect\JournalTitle{Nat. Phys.}}
  \textbf{9}, 541--544 (2013).

\bibitem{melalkia_plug-and-play_2022}
M.~F. Melalkia, T.~Gabbrielli, A.~Petitjean, \emph{et~al.},
  \enquote{Plug-and-play generation of non-{Gaussian} states of light at a
  telecom wavelength,} {\protect\JournalTitle{Optics Express}} \textbf{30},
  45195 (2022).

\bibitem{RaMultiNG2020}
Y.-S. Ra, A.~Dufour, M.~Walschaers, \emph{et~al.}, \enquote{Non--gaussian
  quantum states of a multimode light field,} {\protect\JournalTitle{Nature
  Physics}} \textbf{16}, 144--147 (2020).

\bibitem{Brecht2015}
B.~Brecht, D.~V. Reddy, C.~Silberhorn, and M.~G. Raymer, \enquote{Photon
  temporal modes: A complete framework for quantum information science,}
  {\protect\JournalTitle{Phys. Rev. X}} \textbf{5}, 041017 (2015).

\bibitem{Wu1986}
L.-A. Wu, H.~J. Kimble, J.~L. Hall, and H.~Wu, \enquote{Generation of squeezed
  states by parametric down conversion,} {\protect\JournalTitle{Phys. Rev.
  Lett.}} \textbf{57}, 2520 (1986).

\bibitem{Slusher1985}
R.~E. Slusher, L.~W. Hollberg, B.~Yurke, \emph{et~al.}, \enquote{Observation of
  squeezed states generated by four-wave mixing in an optical cavity,}
  {\protect\JournalTitle{Phys. Rev. Lett.}} \textbf{55}, 2409 (1985).

\bibitem{Barbosa2013}
F.~A.~S. Barbosa, A.~S. Coelho, K.~N. Cassemiro, \emph{et~al.}, \enquote{Beyond
  spectral homodyne detection: Complete quantum measurement of spectral modes
  of light,} {\protect\JournalTitle{Phys. Rev. Lett.}} \textbf{111}, 200402
  (2013).

\bibitem{Barbosa2018}
F.~A.~S. Barbosa, A.~S. Coelho, L.~F. Mu\~noz Mart\'{\i}nez, \emph{et~al.},
  \enquote{Hexapartite entanglement in an above-threshold optical parametric
  oscillator,} {\protect\JournalTitle{Phys. Rev. Lett.}} \textbf{121}, 073601
  (2018).

\bibitem{PRXNice}
L.~Labont\'e, O.~Alibart, V.~D'Auria, \emph{et~al.}, \enquote{Integrated
  photonics for quantum communications and metrology,}
  {\protect\JournalTitle{PRX Quantum}} \textbf{5}, 010101 (2024).

\bibitem{Kogler2024}
R.~Kögler, G.~Rickli, R.~Domeneguetti, \emph{et~al.}, \enquote{Quantum state
  tomography in a third-order integrated optical parametric oscillator,}
  {\protect\JournalTitle{Optics Letters}}  (2024). Received 09 Feb 2024;
  Accepted 09 May 2024; Posted 09 May 2024.

\bibitem{Gouzien2020}
E.~Gouzien, S.~Tanzilli, V.~D'Auria, and G.~Patera, \enquote{Morphing
  supermodes: a full characterization for enabling multimode quantum optics,}
  {\protect\JournalTitle{Phys. Rev. Lett.}} \textbf{125}, 103601 (2020).

\bibitem{Gouzien2023}
E.~Gouzien, L.~Labont\'e, J.~Etesse, \emph{et~al.}, \enquote{Hidden and
  detectable squeezing from microresonators,} {\protect\JournalTitle{Phys. Rev.
  Res.}} \textbf{5}, 023178 (2023).

\bibitem{Miller2023}
J.~L. Miller, \enquote{Frequency-dependent squeezing pushes ligo sensitivity to
  new records,} {\protect\JournalTitle{Physics Today}} \textbf{2023}, 1116a
  (2023).

\bibitem{Kimble2001}
H.~J. Kimble, Y.~Levin, A.~B. Matsko, \emph{et~al.}, \enquote{Conversion of
  conventional gravitational-wave interferometers into quantum nondemolition
  interferometers by modifying their input andÕor output optics,}
  {\protect\JournalTitle{Phys. Rev. D}} \textbf{65}, 022002 (2001).

\bibitem{McCuller2020}
L.~McCuller, C.~Whittle, D.~Ganapathy, \emph{et~al.},
  \enquote{Frequency-dependent squeezing for advanced ligo,}
  {\protect\JournalTitle{Phys. Rev. Lett.}} \textbf{124}, 171102 (2020).

\bibitem{Yuen1980}
H.~P. Yuen and J.~H. Shapiro, \enquote{Optical communication with two-photon
  coherent states- {Part III}: Quantum measurements realizable with
  photoemissive detectors,} {\protect\JournalTitle{IEEE Trans. Inform. Theory}}
  \textbf{26}, 78 (1980).

\bibitem{Yuen1983}
H.~P. Yuen and V.~W.~S. Chan, \enquote{Noise in homodyne and heterodyne
  detection,} {\protect\JournalTitle{Opt. Lett.}} \textbf{8}, 177 (1983). {\bf
  8}, 345 (1983).

\bibitem{BarbosaPRA2013}
F.~A.~S. Barbosa, A.~S. Coelho, K.~N. Cassemiro, \emph{et~al.},
  \enquote{Quantum state reconstruction of spectral field modes: Homodyne and
  resonator detection schemes,} {\protect\JournalTitle{Phys. Rev. A}}
  \textbf{88}, 052113 (2013).

\bibitem{FabreReview}
C.~Fabre and N.~Treps, \enquote{Modes and states in quantum optics,}
  {\protect\JournalTitle{Rev. Mod. Phys.}} \textbf{92}, 035005 (2020).

\bibitem{Pfister2019}
O.~Pfister, \enquote{Continuous-variable quantum computing in the quantum
  optical frequency comb,} {\protect\JournalTitle{Journal of Physics B: Atomic,
  Molecular and Optical Physics}} \textbf{53}, 012001 (2020).

\bibitem{Patera2010}
G.~Patera, N.~Treps, C.~Fabre, and G.~J. de~Valc\'{a}rcel, \enquote{Quantum
  theory of synchronously pumped type i optical parametric oscillators:
  generation of multiple, squeezed frequency combs below threshold,}
  {\protect\JournalTitle{Eur. Phys. J. D}} \textbf{56}, 123 (2010).

\bibitem{Chembo2016}
Y.~K. Chembo, \enquote{Quantum dynamics of kerr optical frequency combs below
  and above threshold: Spontaneous four-wave mixing, entanglement, and squeezed
  states of light,} {\protect\JournalTitle{Phys. Rev. A}} \textbf{93}, 033820
  (2016).

\bibitem{Christ2011}
A.~Christ, K.~Laiho, A.~Eckstein, \emph{et~al.}, \enquote{Probing multimode
  squeezing with correlation functions,} {\protect\JournalTitle{New J. Phys.}}
  \textbf{13}, 033027 (2011).

\bibitem{GardinerZoller}
C.~W. Gardiner and P.~Zoller, \emph{Quantum Noise} (Springer, 1991).

\bibitem{Ferraro_book}
A.~Ferraro, S.~Olivares, and M.~G.~A. Paris, \emph{Gaussian states in
  continuous variable quantum information} (Bibliopolis, Napoli, 2005).

\bibitem{Dauria2010}
V.~D'Auria, S.~Fornaro, A.~Porzio, \emph{et~al.}, \enquote{Full
  characterization of gaussian bipartite entangled states by a single homodyne
  detector,} {\protect\JournalTitle{Phys. Rev. Lett.}} \textbf{102}, 020502
  (2009).

\bibitem{Laurat2005}
J.~Laurat, L.~Longchambon, C.~Fabre, and T.~Coudreau, \enquote{Experimental
  investigation of amplitude and phase quantum correlations in a type ii
  optical parametric oscillator above threshold: from nondegenerate to
  degenerate operation,} {\protect\JournalTitle{Opt. Lett.}} \textbf{30}, 1177
  (2005).

\bibitem{Guidry2023}
M.~A. Guidry, D.~M. Lukin, K.~Y. Yang, and J.~Vu{\v{c}}kovi{\'c},
  \enquote{Multimode squeezing in soliton crystal microcombs,}
  {\protect\JournalTitle{Optica}} \textbf{10}, 694--701 (2023).

\bibitem{Fabre1989}
C.~Fabre, E.~Giacobino, A.~Heidmann, and S.~Reynaud, \enquote{Noise
  characteristics of a non-degenerate optical parametric oscillator ---
  application to quantum noise reduction,} {\protect\JournalTitle{J. Phys.
  (Paris)}} \textbf{50} (1989).

\bibitem{Roslund2014}
J.~Roslund, R.~{Medeiros de Ara\'ujo}, S.~Jiang, \emph{et~al.},
  \enquote{Wavelength-multiplexed quantum networks with ultrafast frequency
  combs,} {\protect\JournalTitle{Nat. Photon.}} \textbf{8}, 109 (2014).

\bibitem{Fabre1990}
C.~Fabre, E.~Giacobino, A.~Heidmann, \emph{et~al.}, \enquote{Squeezing in
  detuned degenerate optical parametric oscillators,}
  {\protect\JournalTitle{Quantum Optics}} \textbf{2}, 159 (1990).

\bibitem{Solimeno2002}
A.~Porzio, C.~Altucci, P.~Aniello, \emph{et~al.}, \enquote{Resonances and
  spectral properties of detuned opo pumped by fluctuating sources,}
  {\protect\JournalTitle{Appl.Phys. B}} \textbf{75}, 655--665 (2002).

\bibitem{Fabre1994}
C.~Fabre, M.~Pinard, S.~Bourzeix, \emph{et~al.}, \enquote{Quantum-noise
  reduction using a cavity with a movable mirror,} {\protect\JournalTitle{Phys.
  Rev. A}} \textbf{49}, 1337 (1994).

\bibitem{Mancini1994}
S.~Mancini and P.~Tombesi, \enquote{Quantum noise reduction by radiation
  pressure,} {\protect\JournalTitle{Phys. Rev. A}} \textbf{49}, 4055 (1994).

\bibitem{Junker2022}
J.~Junker, D.~Wilken, N.~Johny, \emph{et~al.}, \enquote{Frequency-dependent
  squeezing from a detuned squeezer,} {\protect\JournalTitle{Phys. Rev. Lett.}}
  \textbf{129}, 033602 (2022).

\bibitem{Yanagimoto2022}
R.~Yanagimoto, E.~Ng, A.~Yamamura, \emph{et~al.}, \enquote{Onset of
  non-gaussian quantum physics in pulsed squeezing with mesoscopic fields,}
  {\protect\JournalTitle{Optica}} \textbf{9}, 379 (2022).

\bibitem{Jankowski2024}
M.~Jankowski, R.~Yanagimoto, E.~Ng, \emph{et~al.}, \enquote{Ultrafast
  second-order nonlinear photonics—from classical physics to non-gaussian
  quantum dynamics,} {\protect\JournalTitle{Advances in Optics and Photonics}}
  (2024).

\bibitem{Bachor2019}
H.-A. Bachor and T.~C. Ralph, \emph{A Guide to Experiments in Quantum Optics}
  (Wiley-VCH, 2019), 3rd ed.

\bibitem{Buchmann2016}
L.~F. Buchmann, S.~Schreppler, J.~Kohler, \emph{et~al.}, \enquote{Complex
  squeezing and force measurement beyond the standard quantum limit,}
  {\protect\JournalTitle{Phys. Rev. Lett.}} \textbf{117}, 030801 (2016).

\bibitem{Li2013}
C.-K. Li, R.~Roberts, and X.~Yin, \enquote{Decomposition of unitary matrices
  and quantum gates,}  (2013).

\bibitem{Hu2021}
Y.~Hu, M.~Yu, D.~Zhu, \emph{et~al.}, \enquote{On-chip electro-optic frequency
  shifters and beam splitters,} {\protect\JournalTitle{Nature}} \textbf{599},
  587--593 (2021).

\bibitem{Lu2018}
H.-H. Lu, J.~M. Lukens, N.~A. Peters, \emph{et~al.}, \enquote{Electro-optic
  frequency beam splitters and tritters for high-fidelity photonic quantum
  information processing,} {\protect\JournalTitle{Phys. Rev. Lett.}}
  \textbf{120}, 030502 (2018).

\bibitem{Reck1994}
M.~Reck, A.~Zeilinger, H.~J. Bernstein, and P.~Bertani, \enquote{Experimental
  realization of any discrete unitary operator,} {\protect\JournalTitle{Phys.
  Rev. Lett.}} \textbf{73}, 58--61 (1994).

\bibitem{Clements2016}
W.~R. Clements, P.~C. Humphreys, B.~J. Metcalf, \emph{et~al.}, \enquote{Optimal
  design for universal multiport interferometers,}
  {\protect\JournalTitle{Optica}} \textbf{3}, 1460--1465 (2016).

\bibitem{Petersen2011}
I.~R. Petersen, \enquote{Cascade cavity realization for a class of complex
  transfer functions arising in coherent quantum feedback control,}
  {\protect\JournalTitle{Automatica}} \textbf{47}, 1757--1763 (2011).

\bibitem{Nurdin2009}
H.~I. Nurdin, M.~R. James, and A.~C. Doherty, \enquote{Network synthesis of
  linear dynamical quantum stochastic systems,} {\protect\JournalTitle{{SIAM}
  Journal on Control and Optimization}} \textbf{48}, 2686--2718 (2009).

\bibitem{Dioum2024}
B.~Dioum, V.~D'Auria, and G.~Patera, \enquote{Hidden quantum correlations in
  cavity-based quantum optics,}  (2024). Draft paper.

\bibitem{Bensemhoun2024}
A.~Bensemhoun, C.~Gonzalez-Arciniegas, O.~Pfister, \emph{et~al.},
  \enquote{Multipartite entanglement in bright frequency combs out of
  microresonators,} {\protect\JournalTitle{Physics Letters A}} \textbf{493},
  129272 (2024).

\end{thebibliography}
%\bibliographyfullrefs{biblio,Pfister}

\end{document}